\documentclass[review]{elsarticle}

\usepackage{lineno,hyperref}
\modulolinenumbers[5]

\journal{Journal of \LaTeX\ Templates}

\usepackage{here}
\usepackage{moreverb}
\usepackage{soul}
\usepackage{amsmath}
\usepackage{amssymb}
\usepackage{algorithm}
\usepackage{algpseudocode}
\usepackage{graphicx}
\usepackage{subfigure}
\usepackage{array}

\begin{document}

\begin{frontmatter}


\title{Proportional Fair Scheduling Using Water-Filling Technique for SC-FDMA Based D2D Communication}

\author[UofG]{Syed Tariq Shah \corref{mycorrespondingauthor}}
\ead{syedtariq.shah@glasgow.ac.uk}

\author[SKKU]{Jaheon Gu }
\ead{mageboy@skku.edu}

\author[AUS]{Syed Faraz Hasan}
\ead{faraz.hasan@une.edu.au}

\author[SKKU]{Min Young Chung\corref{mycorrespondingauthor}}
\cortext[mycorrespondingauthor]{Corresponding author}
\ead{mychung@skku.edu}

\address[UofG]{James Watt School of Engineering, University of Glasgow, Glasgow, G12 8QQ, UK}
\address[SKKU]{Department of Electrical and Computer Engineering, Sungkyunkwan University, Suwon, South Korea}
\address[AUS]{Directorate of Research and Enterprise, University of New England, Australia}

\begin{abstract}
The resource allocation in SC-FDMA is constrained by the condition that multiple subchannels should be allocated to a single user only if they are adjacent. Therefore, the scheduling scheme of a D2D-cellular system that uses SC-FDMA must also conform to the so-called adjacency constraint. This paper proposes a heuristic algorithm with low computational complexity that applies proportional fair (PF) scheduling in the D2D-cellular system. The proposed algorithm consists of two main phases: i) subchannel allocation and ii) adjustment of data rates, which are executed for both CUEs and DUEs. In the subchannel allocation phase for CUEs (or D2D pairs), the users' data rates are maximized via optimal power allocation to frequency-contiguous subchannels. In the second phase, a PF scheduling problem is solved to decide the modulation and coding scheme (MCS) of both CUEs and D2D pairs. Both phases of the proposed algorithm benefit from the Water-Filling (WF) technique. The simulation results suggest that the proposed scheme performs similar to optimal PF scheduling from the perspective of users' data rate and their logarithmic sum. An additional benefit of the proposed scheme is its low computational overhead.
\end{abstract}


\end{frontmatter}


\section{Introduction}
\label{intro}
With the increased use of high-quality multimedia, the volume of the mobile service traffic generated from wireless devices has increased considerably \cite{{ref1},{ref2},{ref3}}. 
In the conventional cellular communication system, a centralized coordinator, an evolved NodeB (eNB), relays data traffic between several cellular user equipment (CUEs). 
The central eNB can be severely bottlenecked as it relays massive traffic volumes to and from CUEs.
In order to offload the traffic going through eNB, the 3rd Generation Partnership Project (3GPP) has introduced Device-to-Device (D2D) communications for Long Term Evolution Advanced (LTE-A) system \cite{ref4}.
D2D communication is expected to reduce the traffic flowing through eNB by enabling two users to share data on direct links.
Such a user is called D2D user equipment (DUE), which pairs up with another DUE to send and receive data without involving the eNB. \textbf{Moreover, the D2D in cooperative communication scenarios also leads to better end-to-end delivery and network energy efficiency \cite{Ali}.} 

\begin{figure}
 \centering
 \subfigure[Case 1: DUEs using UL frequencies.]{
     \label{F1a}
     \includegraphics[width=5.4cm]{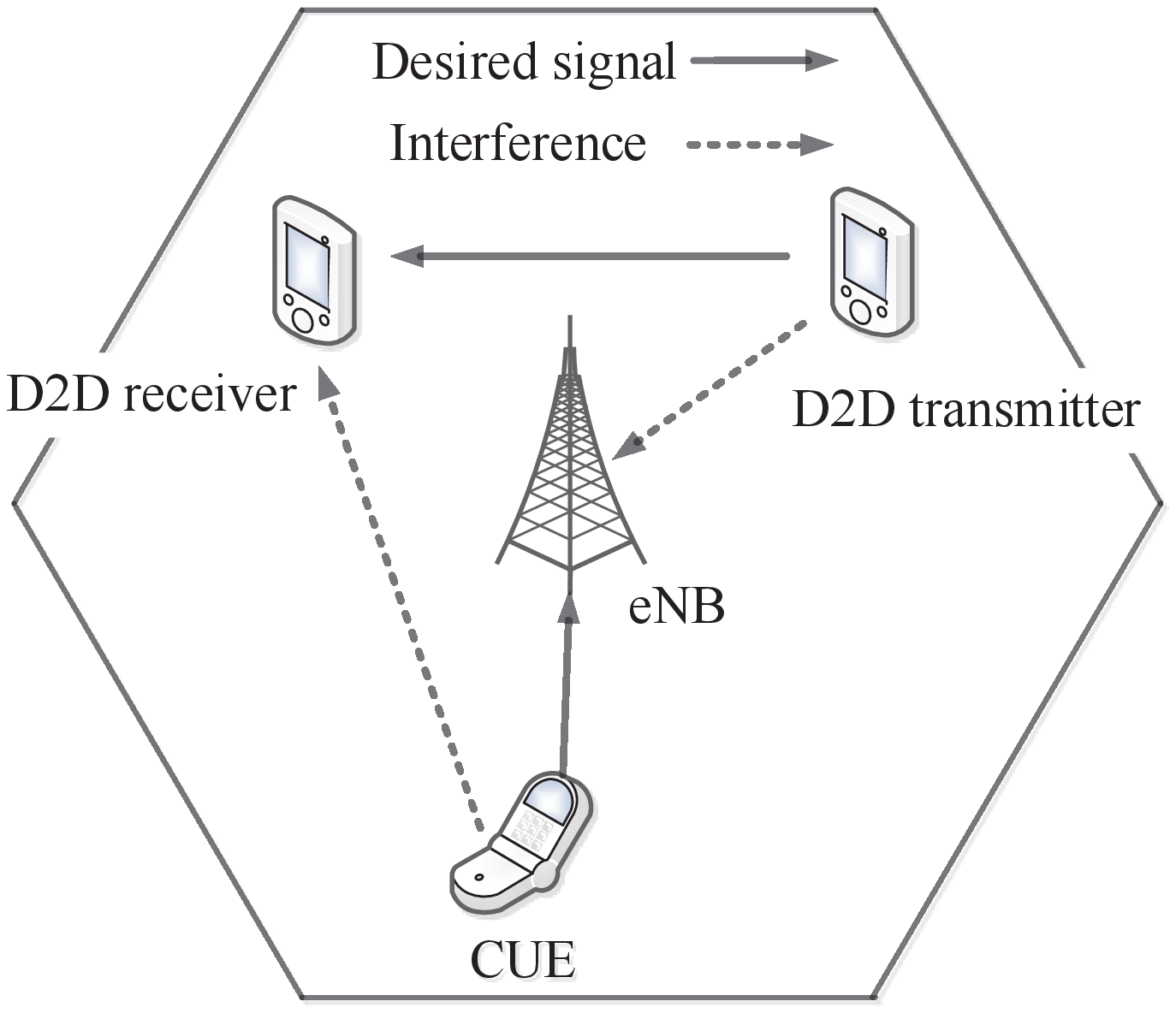}}
 \subfigure[Case 2: DUEs using DL frequencies.]{
        \label{F1b}
        \includegraphics[width=5.4cm]{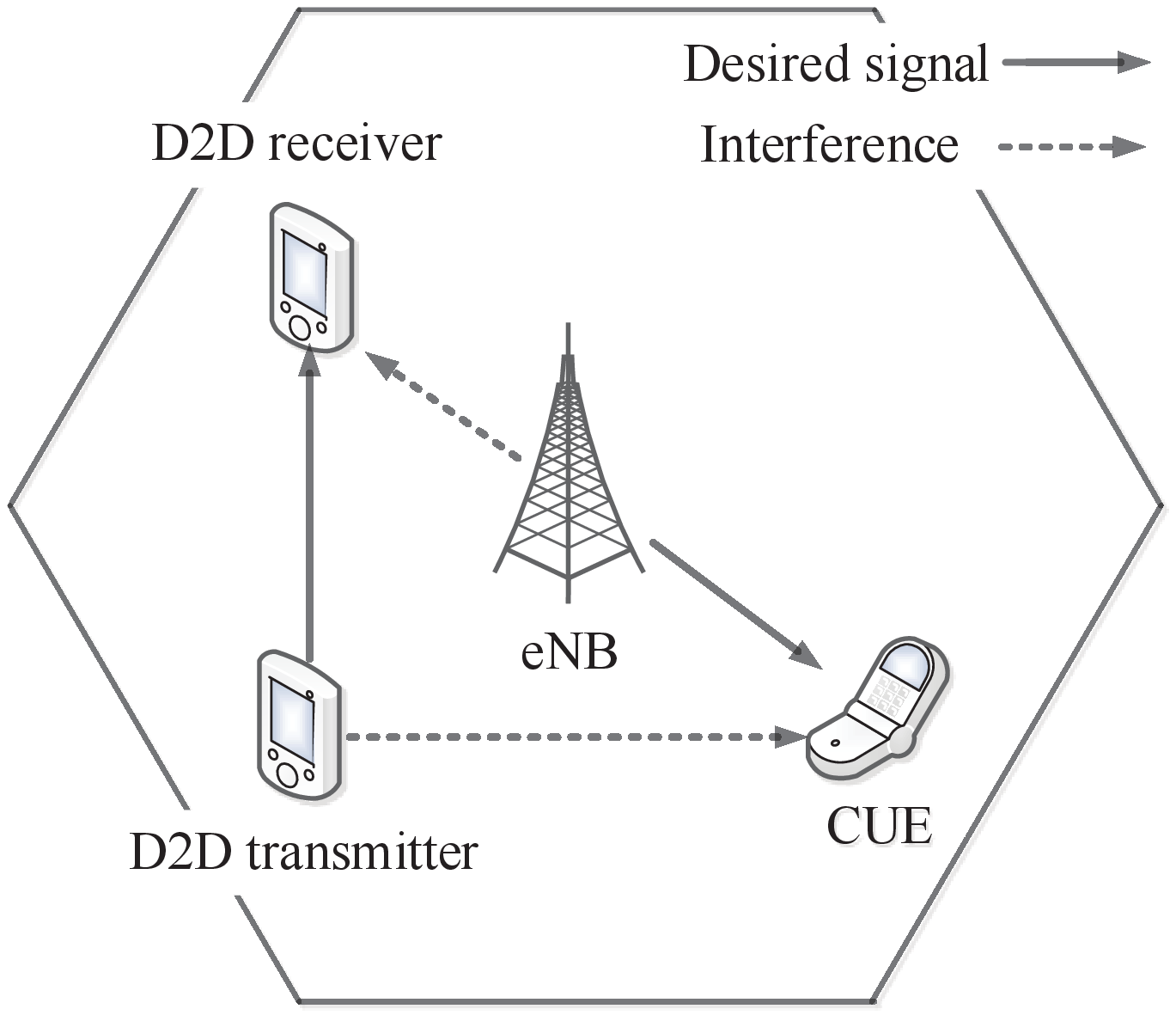}}
\caption{Interference signals generating in UL/DL frequencies of the cellular network.}
\label{F1}
\end{figure}

The underlying version of D2D communications reuses the uplink (UL) and/or downlink (DL) frequency bands of the cellular networks \cite{ref5}.
When DUEs reuse radio resources primarily meant for CUEs, heavy interference between DUE and CUE (and also with eNB) may be observed.
Case 1 in Fig. \ref{F1} shows a typical scenario when DUEs use the UL frequency band of the cellular networks. 
In this scenario, D2D transmitters and CUEs can cause considerable interference to eNB and other devices using the same resource.
Case 2 in Fig. \ref{F1} represents a scenario in which DUEs reuse the DL frequency band of the cellular networks.
In this case, the D2D transmitter and eNB may cause interference to CUEs and D2D receivers, respectively.
Since the traffic loads in a real-life cellular system are asymmetric, it is expected that the DUE prefers to use the UL frequency band than the DL frequency band \cite{{ref6},{GFodor},{Tariq_WPC}}.
This paper also considers an environment where DUEs reuse the UL frequency band of LTE-A.

The LTE-A system divides the available frequency band into several subchannels, each comprising 12 subcarriers.
An eNB allocates the available subchannels to the existing CUEs for communication. 
The users, being much higher in number, share the available subchannels among themselves.
LTE-A uses orthogonal frequency division multiple access (OFDMA) on the DL and single-carrier frequency division multiple access (SC-FDMA) on the UL to share the spectrum. 
OFDMA has high spectral efficiency and is not affected by multi-path fading \cite{ref10}.
While these advantages increase data rates and the overall system capacity, the main drawback of OFDMA is a high peak-to-average power ratio (PAPR) \cite{ref11}. \textbf{From the perspective of non-orthogonal multiple access (NOMA), it is shown in \cite{MKA} that a many-to-many matching game and a DC programming-based sub-channel and power allocation scheme for D2D-enabled NOMA cellular communications can enhance the spectral efficiency, network connectivity, and fairness of the networks.}

The use of SC-FDMA on the UL is primarily motivated by the fact that it reduces PAPR.
SC-FDMA divides the transmit power of user equipment (UE) into multiple subchannels by spreading the OFDM symbols. 
This spreading of symbols reduces PAPR and brings a number of benefits, including transmit power efficiency \cite{{ref12},{ref13}}. SC-FDMA uses two subchannel mapping techniques: one for localized SC-FDMA (LFDMA) and the second for distributed SC-FDMA (DFDMA).
Subchannels assigned to a user in the frequency domain need not be contiguous in DFDMA.
In DFDMA, the SC-FDMA signals received from distributed resources are complex to decode, and therefore it yields poor spectral efficiency and data rates compared to LFDMA \cite{ref14}.
Therefore, LFDMA is intuitively a preferred choice in this paper.
LFDMA has exclusivity, and adjacency constraints for resource allocation \cite{ref13}.
The exclusivity constraint is that a subchannel could be assigned to a single CUE,
which means that the CUE should use a subchannel that is not (re)used by any other CUE.
The adjacency constraint means that multiple subchannels must be contiguous regarding frequencies if the subchannels are assigned to a single CUE. 

In addition to multiple access techniques (SC-FDMA and OFDM), the performance of a cellular network is considerably affected by radio resource scheduling.
\cite{JGu,JLee,STShah,STShah_JWCN}.
Different scheduling schemes are readily available, focusing on providing a fair chance for all users to use the cellular network.
Among the available scheduling schemes, the proportional fair (PF) scheme promises a fair trade-off between the average achievable user data rates and users' fairness \cite{JGu:TWireless}.
In order to be used in conjunction with SC-FDMA, PF scheduling needs to consider the exclusivity and adjacency constraints mentioned above.
Several related studies have examined the scheduling issues in an SC-FDMA system \cite{JGu:TWireless,{ref17},{ref18}}.
These studies either assign subchannels to only CUEs through the LFDMA or subchannels to both CUEs and DUEs without considering the constraints of LFDMA.
Moreover, the constraints in SC-FDMA increase the computational complexity of PF scheduling. A computationally expensive system is, of course, undesirable.
To the best of our knowledge, no previous work has applied PF scheduling in conjunction with LFDMA for D2D communications.

\subsection*{Paper Objectives and Contributions}

This paper proposes a heuristic algorithm for proportional fair resource scheduling in SC-FDMA-based D2D-Cellular System that complies with 3GPP LTE-A standard \cite{ref19,ref20}.
One of the major contributions of the proposed scheme is that it applies the Water Filling (WF) technique for subchannel allocation.
The existing works have used the WF technique mainly for transmit power control of CUEs \cite{HWang, EBaccarelli}.
The NP-hard problem of resource scheduling can be solved in polynomial time by efficiently applying the WF technique, where the resources are allocated to users in a step-by-step manner.
The advantages of the proposed WF-based resource scheduler are two folds; first, it has significantly lower computational complexity than an optimal PF scheduler.
Second, for both CUEs and DUEs, the performance of the proposed scheme in terms of achievable user data rates is very close to that optimal PF scheduling scheme.
Since the proposed algorithm considers two constraints, it operates in two phases: 
i) subchannel allocation phase and ii) adjustment of data rates phase.
In the subchannel allocation phase, a WF-based approach is used to allocate the subchannels to both DUEs and CUEs under the adjacency constraint.
In the second phase (i.e., data rate adjustment phase), the proposed scheme decides the MCS of both CUEs and D2D pairs to maximize their logarithmic sum of average data rates pseudo-optimally.
These two phases can be repeated (i) for $M$ times or (ii) until no further change is observed in allocated resources to minimize interference.
Our simulation analysis shows that the proposed scheme can reduce the computational complexity of the PF scheduling and can drastically increase the logarithmic sum of average user data rates.

The remainder of this paper is organized as follows.
Section 2 briefly describes the previous works on PF scheduling for conventional cellular systems. Any scheduling algorithm for LFDMA-based D2D communication is not available to the best of the author's knowledge.
The considered system model and optimal PF scheduling problem are illustrated in Section 3. The proposed heuristic algorithm for SC-FDMA-based PF scheduling is introduced in Section 4.
The performance of optimal PF scheduling and the proposed heuristic algorithm is discussed in Section 5. This paper is concluded in Section 6.

\section{Background}

The present state of the art in PF scheduling addresses the SC-FDMA-based conventional cellular systems where D2D pairs do not exist.
Before proposing PF scheduling in SC-FDMA-based Cellular-D2D systems, this section briefly outlines how SC-FDMA affects scheduling in the conventional cellular system excluding D2D. We use this knowledge in Section 3 to discuss scheduling in D2D communication.

In order to allocate adjacent resources to UEs, 
an optimal algorithm is proposed to maximize the logarithmic sum-rate in SC-FDMA-based systems \cite{ref15}.
It finds every subchannel allocation pattern and calculates the PF metric, which is the ratio of the instantaneous data rate to the average data rate for all the patterns.
Then, it chooses and allocates the subchannel set, maximizing the sum of PF metrics.
This algorithm has high computing complexity since it considers every feasible subchannel for every UE.
Therefore, the proposed algorithm could not be adopted practically due to a high computational cost.

For reducing the computing complexity of the PF scheduling, a heuristic algorithm has been presented in \cite{ref16}.
It results in the high performance of UEs and offers flexibility by introducing ranking threshold $T_r$.
For example, if $T_r=1$, a subchannel would be allocated to the UE with the best PF metric on the subchannel.
If $T_r=2$, the subchannel can be allocated to the UE with either 1st or 2nd highest PF metric.
In general, if $T_r=\delta$, the subchannel can be allocated to the UE with either 1st, 2nd, $\cdots$, or $\delta$th highest PF metric.
This enhances the flexibility of the subchannel allocation compared to the existing algorithm. 
On the other hand, the algorithm increases its complexity versus increasing $T_r$.

Another scheduling method has been illustrated in \cite{ref17}.
The method allocates a subchannel to a UE based on the marginal PF metric,
which is the potential PF metric assuming that the UE is already assigned to subchannels.
The subchannel at which (i) the UE's marginal PF metric is highest and (ii) UE's marginal PF metric shows the highest difference from that of the others is allocated to the UE.
Since the algorithm in \cite{ref17} takes both the PF metric difference and the marginal PF metrics into account, the fairness of SC-FDMA could be improved.

For alleviating the computing complexity of the PF scheduling, a coherence sub-band-based resource allocation (CSRA) algorithm was introduced for SC-FDMA in \cite{ref18}.
For assigning subchannels to UEs following the constraints, CSRA comprises two steps, namely $enclosure$ and $contributing$.
In $enclosure$ step, the entire frequency band is split to multiple subbands based on the channel coherence.
Then, eNB calculates the PF metric on a subband, which is the sum of PF metric values on every subchannel included in the subband.
A subband may include one or more subchannels according to channel coherence and fading characteristics.
Based on the calculated value, the UE with the best PF metric on a subband would be assigned to the subchannel with the best PF metric among subchannels in the subband.
In the $contributing$ step, the remaining subchannels are assigned to UEs based on the maximum contribution.
The UE highly contributing to the objective function would be chosen for a particular subchannel, and then assigned to the subchannel.
The $contributing$ step is iterated until every subchannel is assigned.
CSRA could nearly touch the optimum performance in the perspective of a logarithmic utility function with lower complexity.
Despite these benefits, the scheme in \cite{ref17} cannot allocate resources for D2D communications.
This is because interference between cellular and D2D users is ignored.

Our previous work also deals with the scheduling in cellular networks without D2D communications \cite{ett:jgu}.
Our previous work presented a water-filling-based optimized solution for PF scheduling in the legacy cellular network (without D2D communications).
Furthermore, the work in \cite{ett:jgu} could not directly be applied to D2D communications-enabled networks because of the interference between CUEs and DUEs.
In order to reflect the interference in a two-tier network (cellular and D2D), this paper proposes a novel heuristic algorithm which iteratively executes the water-filling technique for resource allocation for both cellular and D2D users in the network.

\section{Proposed Heuristic PF Scheduling}

\subsection{System Model}

In contrast with the existing literature, we consider a system with CUE $i_c$ ($=1, 2, \cdots, N_{C}$) and D2D pairs $i_d$ ($=1,2,\cdots,N_{D}$), where $N_{C}$ and $N_{D}$ are the total number of CUEs and D2D pairs, respectively.
The D2D Tx and its paired Rx are deployed in a cell covered by a single eNB.
Subchannel $k$ ($= 1, 2, \cdots, K$) is assigned to a CUE and a D2D pair by the scheduler $S$ considering the exclusivity constraint.
According to \cite{JGu:TWireless}, for any D2D system using $K$ subchannels, the scheduler $P$ is said to be proportionally fair only when it satisfies the:
\begin{eqnarray}
P =
\begin{cases} \arg \max_{S} \left\{ \sum_{i_{c} \in U_{c}^{S}} \ln \left[ \sum_{k=1}^{K} a_{i_c, k}^{(S)} r_{i_c, k}^{(S)} \right] \right.\\
\! + \! \left. \sum_{i_{d} \in U_{d}^{S}} \ln \left[ \sum_{k=1}^{K} a_{i_d, k}^{(S)} r_{i_d, k}^{(S)} \right]\right\}, \quad \textrm{for $T=1$,}\!\!\!\!\!\!\\
\arg \max_{S} \left\{ \sum_{i_{c} \in U_{c}^{S}} \ln \left[1 \! + \!\frac{\sum_{k=1}^{K} a_{i_c, k}^{(S)} r_{i_c, k}^{(S)}}{(T-1) {\bar R}_{i_c}} \right]\right. \\
\! + \! \left. \sum_{i_{d} \in U_{d}^{S}} \ln \!\!\left[1 \!+ \! \frac{\sum_{k=1}^{K} a_{i_d, k}^{(S)} r_{i_d, k}^{(S)}}{(T-1) {\bar R}_{i_d}} \right]\right\}, \textrm{for $T\geq2$,}\!\!\!\!\!\! \label{approximatedPF} \end{cases}
\end{eqnarray}
where $P$ is the PF scheduling and $S$ is any other scheduling and $T$ is the averaging window size.
$U_{c}^{P}$ and $U_{c}^{S}$ are the sets of users in the cellular network selected by the schedulers $P$ and $S$, respectively, while $U_{d}^{P}$ and $U_{d}^{S}$ are the sets of D2D pairs in the D2D network selected by the schedulers $P$ and $S$, respectively.
$a_{i_c,k}^{(S)}$ and $a_{i_d,k}^{(S)}$ are the binary assignment variables pointing if the subchannel $k$ is assigned to the user $i_c$ and $i_d$ by the scheduler $S$, respectively. Likewise, the
$r_{i_{c(\textrm{or }d)},k}^{(S)}$ is the achievable instantaneous data rate of user $i_{c}$ (or $i_{d}$) when the scheduler $S$ assigns the users on subchannel $k$, under $k \in C_{i_{c}}$ (or $k \in C_{i_{d}}$).
The average of user $i_c$'s and $i_d$'s data rates achieved by scheduling $S$ are represented by $R_{i_c}^{(S)}$ and $R_{i_d}^{(S)}$, respectively.

It has been pointed out that the major drawback in PF scheduling using OFDMA is that its computational complexity increases as the number of users and subchannels in the network increases.
Precisely speaking, the PF scheduling for the D2D system would have a computing complexity of $O((N_C \times N_D)^K)$, which exponentially rises with $K$ \cite{JGu:TWireless}.
High computing complexity makes implementing optimal PF scheduling in practical networks impossible.
An LTE-A system, which has a large number of subchannels (at least 25 subchannels) and short scheduling intervals (at least one millisecond), can reflect the difficulty in applying optimal scheduling in practical systems.
This computational complexity becomes much higher when the adjacency constraint is additionally considered (for example, in an SC-FDMA system). \textbf{Table \ref{tb:TN} contains the list of frequently used notions in this paper.}

\begin{table}[]
\caption{Table of Notations}
\centering
\label{tb:TN}
\begin{tabular}{|l|l|l|}
\hline
\textbf{S. No.} & \textbf{Notation} & \textbf{Explanation} \\ \hline
1 & $T_r$ & Ranking threshold \\ \hline
2 & $N_{C}$ & The total number of CUEs \\ \hline
3 & $N_{D}$ & The total number of D2D pairs \\ \hline
4 & $K$ & The total number subchannels \\ \hline
5 & $U_{c}^{P}$ and $U_{c}^{S}$ & \begin{tabular}[c]{@{}l@{}}The sets of users in the cellular network selected \\ by the schedulers  $P$ and $S$, respectively\end{tabular} \\ \hline
6 & $U_{d}^{P}$ and $U_{d}^{S}$ & \begin{tabular}[c]{@{}l@{}}The sets of D2D pairs in the D2D network selected\\  by the schedulers $P$ and $S$, respectively\end{tabular} \\ \hline
7 & $a_{i_c,k}^{(S)}$ and $a_{i_d,k}^{(S)}$ & \begin{tabular}[c]{@{}l@{}}The binary assignment variables pointing if the \\ subchannel $k$ is assigned to the user $i_c$ and\\  $i_d$ by the scheduler $S$, respectively\end{tabular} \\ \hline
8 & $r_{i_{c(\textrm{or }d)},k}^{(S)}$ & \begin{tabular}[c]{@{}l@{}}The achievable instantaneous data rate of user $i_{c}$ (or $i_{d}$)\\  when the scheduler $S$ assigns the users on subchannel $k$\end{tabular} \\ \hline
9 & $R_{i_c}^{(S)}$ and $R_{i_d}^{(S)}$ & \begin{tabular}[c]{@{}l@{}}The respective average of user $i_c$'s and $i_d$'s data rates\\  achieved by scheduling $S$.\end{tabular} \\ \hline
10 & $\mathbf{K_{i_c}}$ and $\mathbf{K_{i_d}}$ & The sets of subchannels allocated to $i_c$ and $i_d$, respectively \\ \hline
11 & $r_{i_c,k}$ & \begin{tabular}[c]{@{}l@{}}The instantaneous achievable data rate of CUE $i_c$ on \\ subchannel $k$  at the current transmission time slot\end{tabular} \\ \hline
\end{tabular}
\end{table}

\subsection{Proposed Scheduling}

Here, we propose a heuristic algorithm that significantly reduces the computational complexity of PF scheduling while offering a reasonable throughput performance.
The proposed algorithm uses the WF technique to find the optimal solution. WF technique finds the optimal solution of the convex optimization problem \cite{LZhuang,PHe}.
Since it is known from the logarithmic function in (\ref{approximatedPF}) that PF scheduling is the convex optimization problem, the WF technique effectively solves this problem, taking into account the additional constraints of SC-FDMA.
In order to take into account the adjacency constraint to the conventional PF scheduling in \cite{JGu:TWireless}, the scheduling $P$ in (\ref{approximatedPF}) should satisfy:
\begin{equation}
\label{eq:CUE_schedulingP}
\exists \, k: \sum_{n = 0} ^{a_{i_c}-1} a_{i_c,k+n} = a_{i_c}
\end{equation}
and
\begin{equation}
\label{eq:DUE_schedulingP}
\exists \, k: \sum_{n = 0} ^{a_{i_d}-1} a_{i_d,k+n} = a_{i_d},
\end{equation}
where $a_{i_c}=\sum_{k\in \mathbf{K_{i_c}}} a_{i_c,k}$ and $a_{i_d}=\sum_{k\in \mathbf{K_{i_d}}} a_{i_d,k}$.
$\mathbf{K_{i_c}}$ and $\mathbf{K_{i_d}}$ are the sets of subchannels allocated to $i_c$ and $i_d$, respectively.
The considered problem is also subject to the following terms:
\begin{eqnarray}
\textrm{subject to: } & \textrm{\ \ \ \ } 0 \le r_{i_c} \le \hat{r}_{i_c}, \ \ \forall i_c; \label{eq:subject1} \\
&\textrm{\ \ \ \ } 0 \le r_{i_d} \le \hat{r}_{i_d}, \ \ \forall i_d; \\
&\sum_{i_c \in \mathbf{U_c^P}} r_{i_c} \le \hat{r}_c, \\
&\sum_{i_d \in \mathbf{U_d^P}} r_{i_d} \le \hat{r}_d, \label{eq:subject2} 
\end{eqnarray}
where $\hat{r}_{i_c}$ and $\hat{r}_{i_d}$ are respectively the maximum achievable data rates of CUE $i_c$ and D2D pair $i_d$ in an SC-FDMA system under the adjacency constraint with the PF scheduling policy.
Similarly, $\hat{r}_c$ is the maximum total achievable data rates of the core network where usually wired backhauls are used.
The $\hat{r}_d$ is the upper limit of the D2D pairs' total data rates, which are theoretically achievable.
Furthermore, $r_{i_c} = \sum_{k\in \mathbf{K}_{i_c}} r_{i_c,k}$ and $r_{i_d} = \sum_{k\in \mathbf{K}_i} r_{i_d,k}$. Here $r_{i_c,k}$ is the instantaneous achievable data rate of CUE $i_c$ on subchannel $k$ at the current transmission time slot, which can be calculated as:
\begin{equation}
\label{eq:rik}
r_{i_c,k}=B \cdot \log_{2} \Big ( {1+\frac{p_{i_c,k}g_{i_c,e_0,k}}{N_0}} \Big ),
\end{equation}
where $B$ is the subchannel bandwidth, $p_{i_c,k}$ is the transmit power of CUE $i_c$ on subchannel $k$, and $g_{i_c,e_0,k}$ is the channel gain between CUE $i_c$ and eNB $e_0$ on subchannel $k$.
The channel gain between devices $x$ and $y$ on subchannel $k$ is represented by $g_{x,y,k}$.

The proposed heuristic algorithm is divided into two phases: subchannel allocation and data rate adjustment that is executed for CUEs and D2D pairs $i_d$. One D2D pair comprises one D2D transmitter $i_d^T$ and one D2D receiver $i_d^R$.
In both phases, for both sets of users, the procedure is repeated for either the $M$ number of times or until no further change is observed in the result.
This repetition removes the interference between legacy cellular communication and D2D communication.
During the subchannel allocation phase for CUEs, the WF technique allocates Shared Control Channels to CUEs under the adjacency constraint.
WF technique also determines the transmission power of CUEs in the same phase.
In the second phase (i.e., data rate adjustment phase), the proposed scheme decides the MCS of both CUEs and D2D pairs to maximize their logarithmic sum of average data rates pseudo-optimally. A more detailed discussion of both phases for CUEs and D2D pairs is provided in the following sections.

\begin{algorithm} 
\caption{Proposed heuristic PF scheduling (PHPFS) algorithm.}
\label{A1} 
\begin{algorithmic}[1]
\State $t \gets 1$
\While {$t \le M$ and $\Gamma_{c,t-1} \ne \Gamma_{c,t-2}$ and $\Gamma_{d,t-1} \ne \Gamma_{d,t-2}$} 
\Procedure{Subchannel allocation phase for CUEs}{}
\State initialize $\hat{r}_c$, $\mathbf{\Lambda_c}$ (set), $\mathbf{\Gamma_{c,t}}(i_c,k)$ (matrix)
\State set $\mathbf{U_c'} \gets \mathbf{U_c}$ ($=\{1,2,\cdots,N_C\}$)
\While{$\mathbf{U_c'} \ne \phi$}
\State $i_c^{\ast} \gets \arg \min_{i_c \in U_c'} (T-1)\bar{R}_{i_c}$
\For{$k'=1$ to $K$}
\State compute $\{ p_{i_c^{\ast},k}^{(k')} \}$ for $\forall k \in \mathbf{K}$ (refer to (\ref{eq:pik}))
\State compute $\hat{r}_{i_c^{\ast}}^{(k')}$ (refer to (\ref{eq:riastk}))
\EndFor
\State compute $\hat{r}_{i_c^{\ast}}$ and $\mathbf{K}_{i_c^{\ast}}$ (refer to (\ref{eq:maxr}) and (\ref{eq:Ki}))
\State $\mathbf{U_c'} \gets \mathbf{U_c'} \setminus \{i_c^{\ast}\}$
\State $g_{i_c,e_0,k} \gets 0$ \ \ for $\forall i_c \in \mathbf{U_c'}$ and $\forall k \in \mathbf{K}_{i_c^{\ast}}$
\State $\mathbf{\Gamma_{c,t}}(i_c^\ast, k ) \gets 1$ for $\forall k \in \mathbf{K}_{i_c^{\ast}}$
\EndWhile
\EndProcedure 
\Procedure{Data rate adjustment phase for CUEs}{}
\State set $\mathbf{U_c'} \gets \mathbf{U_c}$ ($=\{1,2,\cdots,N\}$)
\While{$\mathbf{U_c'} \ne \phi$}
\State calculate $\{r_{i_c}\}$ for $\forall i_c \in \mathbf{U_c'}$ (refer to (\ref{eq:caseequation}))
\State $\mathbf{\Lambda_c} \gets \{ i_c | r_{i_c} > \hat{r}_{i_c}$, $i_c \in \mathbf{U_c} \} $
\If{$\mathbf{\Lambda_c} \ne \phi$}
\State $r_{i_c} \gets \hat{r}_{i_c}$ for $\forall i_c \in \mathbf{\Lambda_c}$
\Else
\State break;
\EndIf
\State $\mathbf{U'_c} \gets \mathbf{U'_c} \setminus \mathbf{\Lambda_c}$ and $\hat{r}_c \gets \hat{r}_c-\sum_{i_c \in \mathbf{\Lambda_c}} \hat{r}_{i_c}$
\EndWhile
\EndProcedure
\algstore{phase1}
\end{algorithmic} 
\end{algorithm}
\begin{algorithm} 
\begin{algorithmic}[1] 
\algrestore{phase1}
\Procedure{Subchannel allocation phase for D2D pairs}{}
\State initialize $\hat{r}_d$, $\mathbf{\Lambda_{d}}$ (set), $\mathbf{\Gamma_{d,t}}(i_d,k)$ (matrix)
\State set $\mathbf{U_d'} \gets \mathbf{U_d}$ ($=\{1,2,\cdots,N_D\}$)
\While{$\mathbf{U_d'} \ne \phi$}
\State $i_d^{\ast} \gets \arg \min_{i_d \in \mathbf{U_d'}} (T-1)\bar{R}_{i_d}$
\For{$k'=1$ to $K$}
\State compute $\{ p_{i_d^{\ast},k}^{(k')} \}$ for $\forall k \in \mathbf{K}$ (refer to (\ref{eq:pikd2d}))
\State compute $\hat{r}_{i_d^{\ast}}^{(k')}$ (refer to (\ref{eq:riastkd2d}))
\EndFor
\State compute $\hat{r}_{i_d^{\ast}}$ and $\mathbf{K}_{i_d^{\ast}}$ (refer to (\ref{eq:maxrd2d}) and (\ref{eq:Kid2d}))
\State $\mathbf{U_d'} \gets \mathbf{U_d'} \setminus \{i_d^{\ast}\}$
\State $g_{i_d^T,i_d^R,k} \gets 0$ \ \ for $\forall i_d \in U_d'$ and $\forall k \in \mathbf{K}_{i_d^{\ast}}$
\State $\mathbf{\Gamma_{d,t}}(i_d^\ast, k ) \gets 1$ for $\forall k \in \mathbf{K}_{i_d^{\ast}}$
\EndWhile
\EndProcedure 
\Procedure{Data rate adjustment phase for D2D pairs}{}
\State set $\mathbf{U'_d} \gets \mathbf{U_d}$ ($=\{1,2,\cdots,N\}$)
\While{$\mathbf{U_d'} \ne \phi$}
\State calculate $\{r_{i_d}\}$ for $\forall i_d \in \mathbf{U_d'}$ (refer to (\ref{eq:caseequationd2d}))
\State $\mathbf{\Lambda_d} \gets \{ i_d | r_{i_d} > \hat{r}_{i_d}$, $i_d \in \mathbf{U_d} \} $
\If{$\mathbf{\Lambda_d} \ne \phi$}
\State $r_{i_d} \gets \hat{r}_{i_d}$ for $\forall {i_d} \in \mathbf{\Lambda_d}$
\Else
\State break;
\EndIf
\State $\mathbf{U'_d} \gets \mathbf{U'_d} \setminus \mathbf{\Lambda_d}$ and $\hat{r}_d \gets \hat{r}_d-\sum_{i_d \in \mathbf{\Lambda_d}} \hat{r}_{i_d}$
\EndWhile
\EndProcedure 
\State $t \gets t+1$
\EndWhile
\end{algorithmic} 
\end{algorithm}

\section{Resource Allocation and Data Rate Adjustment}

\begin{figure}
\centering
\includegraphics[width=8.3cm]{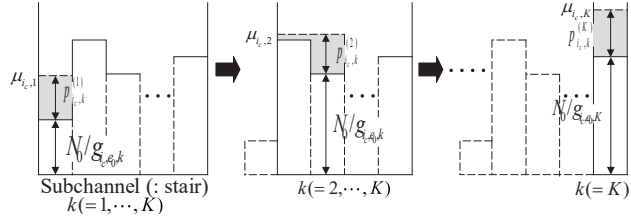}
\caption{The subchannel allocation (Phase 1) cases of CUE $i_c$.}
\label{fig:WFPower}
\end {figure}

\subsection{Subchannel Allocation Phase for CUEs}
In this phase, we first find all possible subchannel allocations options under the assumption that the CUE $i_c$ transmit power should be assigned to various contiguous subchannels from $k'$ (such that $k'$ $=1,\cdots,K$).
Therefore, a total of $K$ allocations can be considered for each CUE.
An example of subchannel allocation cases 1, 2, $\cdots$, $K$ of user $i$ is shown in Figure \ref{fig:WFPower}. 
Then a WF based approach is enforced to $r_{i_c}$ ($= \sum_{k\in \mathbf{K}_{i_c}} r_{i_c,k}$) in order to find $\mathbf{K}_{i_c}$ that maximizes $r_{i_c}$ under the adjacency constraint. 

While mapping the D2D system into the water-filling paradigm, we consider `$k$' as `the index of the stair'. Similarly, `the transmit power of CUE $i_c$ at subchannel $k$ for $k'$ ($p_{i_c,k}^{(k')}$)' is represented by `the amount of water poured into stair $k$'. Finally, `$N_0 / g_{i_c,e_0,k}$' represents `the step depth of the stair $k$'. Furthermore, the $\mu_{i_c,k'}$ represents the final water level in CUE $i_c$'s water tank for $k'$. 

This procedure of subchannel allocation is carried out for all CUEs.
By utilizing the already acquired information of $\bar{R}_{i_c}$ for all CUEs, the proposed scheme chooses CUE $i_c^{\ast}$ (=$\arg\min_{i_c \in \mathbf{U_c}} (T-1)\bar{R}_{i_c}$).
In order to maximize the data rates of user $i_c^{\ast}$, the subchannel allocation problem can be formulated as: 
\begin{eqnarray}
\max_{\{p_{i_c^{\ast},k}^{(k')}\}} & \sum_{k\in \mathbf{K}} B \cdot \log_{2} \Big ( {1+\frac{p_{i_c^{\ast},k}^{(k')}g_{i_c^{\ast},e_0,k}}{N_0+p_{i_d(k)} \cdot g_{i_d^T(k),i_c^{\ast},k}}} \Big ) \label{eq:water_hatri} \\
\textrm{subject to: } & \!\!\!\!\!\!\!\!\!\!\!\!\!\!\!\!\!\!\!\!\!\!\!\!\!\!\!\!\!\!\!\!\!\!\!\!\!\! p_{i_c^{\ast},k}^{(k')}+\frac{N_0+p_{i_d(k)} \cdot g_{i_d^T(k),i_c^{\ast},k}}{g_{i_c^{\ast},e_0,k}} \nonumber\\
& \!\!\!\!\!\!\!\!\!\!\!\!\ge p_{i_c^{\ast},k-1}^{(k')} + \frac{N_0+p_{i_d(k-1)} \cdot g_{i_d^T(k-1),i_c^{\ast},k-1}}{g_{i_c^{\ast},e_0,k-1}} \nonumber \\
& \ \ \ \ \ \ \ \ \ \textrm{for } \forall k\in \{ k'+1,\cdots,K\}; \label{eq:subject3} \\
& \!\!\!\!\!\!\!\!\!\!\!\!\!\!\!\!\!\!\!\!\!\!\!\!\!\!\!\!\!\!\!\!\!\!\!\!\!\!\!\!\!\!\!\!\!\!\!\!\!\!\!\!\!\!\!\!\!\! \sum_{k \in \mathbf{K}} p_{i_c^{\ast},k}^{(k')} \le \hat{P}_{i_c^{\ast}}, \label{eq:subject4} \\
& \!\!\!\!\!\!\!\!\!\!\!\!\!\!\!\!\!\!\!\!\! p_{i_c^{\ast},k}^{(k')} = 0 \textrm{ for } \forall k \!\! \in \!\! \{ 1,\cdots,k'-1\}, \label{eq:subject5}
\end{eqnarray}
where $p_{i_d(k)} \cdot g_{i_d^T(k),i_c^{\ast},k}$ is the interference caused by CUE $i_c(k)$ to the D2D receiver in the pair $i_d$.
Thus, the interference will be zero for $M=1$ and a non-zero value for $M \ge 2$.
Similarly, $i_d^T(k)$ is the transmitter of D2D pair $i_d$ which already occupies subchannel $k$, $p_{i_d(k)}$ is the transmit power of transmitter of D2D pair $i_d(k)$ and $g_{i_d^T(k),i_c^{\ast},k}$ is the channel gain between D2D transmitter $i_d^T(k)$ and CUE $i_c^{\ast}$.
Moreover, the $\hat{P}_{i_d^{\ast}}$ and $\hat{P}_{i_c^{\ast}}$ are the respective maximum transmit powers of users $i_d^{\ast}$'s and $i_c^{\ast}$'s which are predefined in the system.
The condition in (\ref{eq:subject3}) is designed to satisfy the adjacency constraint of SC-FDMA.
In terms of WF technique, this means that the total height of water level ($p_{i_c^{\ast},k}^{(k')}$) and step depth ($\frac{N_0}{g_{i_c^{\ast},e_0,k}}$) of stair $k$ should be equal to or greater than the overall height of the ($k$-1) stair. 
Likewise, as the water is filled in the stair $k'$ step-by-step, the subchannels (i.e., stairs) can now be contiguously allocated.

In order to calculate the optimal $\{ p_{i_c^{\ast},k}^{(k')} \}$, the proposed scheme utilizes the cap-limited modified WF technique \cite{LZhuang}, according to which the $\{ p_{i_c^{\ast},k}^{(k')} \}$ can be calculated as:
\begin{equation}
p_{i_c^{\ast},k}^{(k')} = \left\{ \begin{array}{ll}
\big[ \mu_{i_c,k'} - \frac{N_0+p_{i_d(k)} \cdot g_{i_d^T(k),i_c^{\ast},k}}{g_{i_c^{\ast},e_0,k}} \big] ^{+}, & \textrm{for } k=k',\\
\big[ \mu_{i_c,k'} - p_{i_c^{\ast},k-1}^{(k')} \\ - \frac{N_0+p_{i_d(k-1)} \cdot g_{i_d^T(k-1),i_c^{\ast},k-1}}{g_{i_c^{\ast},e_0,k-1}} \\ + \frac{N_0+p_{i_d(k)} \cdot g_{i_d^T(k),i_c^{\ast},k}}{g_{i_c^{\ast},e_0,k}} \big] ^{+}, & \\
& \!\!\!\!\!\!\!\!\!\!\!\!\!\!\! \textrm{for } k' < k \le K, \\
0, & \textrm{otherwise,}
\end{array} \right.
\label{eq:pik}
\end{equation}
where $[\phi]^{+} = \max \{ 0, \phi \}$, and $\mu_{i_c,k'}$ is the level of water for CUEs in case $k'$ for which (\ref{eq:subject4}) is satisfied with equality.
In addition, the $\hat{r}_{i_c^{\ast}}^{(k')}$ can also be computed as:
\begin{equation}
\hat{r}_{i_c^{\ast}}^{(k')} = \sum_{k\in \mathbf{K}} B \cdot \log_{2} \Big ( {1+\frac{p_{i_c^{\ast},k}^{(k')}g_{i_c^{\ast},e_0,k}}{N_0+p_{i_d(k)} \cdot g_{i_d^T(k),i_c^{\ast},k}}} \Big ).
\label{eq:riastk}
\end{equation}
Then, $\hat{r}_{i_c^{\ast}}$ is decided by
\begin{equation}
\hat{r}_{i_c^{\ast}} = \max_{\{k'\}} \hat{r}_{i_c^{\ast}}^{(k')}
\label{eq:maxr}
\end{equation}
for the case $\hat{k}'$ ($= \arg \max_{\{k'\}} \hat{r}_{i_c^{\ast}}^{(k')}$).
After this, we determine the set $\mathbf{K}_{i_c^{\ast}}$ of subchannels, which will be actually allocated to CUE $i_c^{\ast}$ for $\hat{k}'$.
From (\ref{eq:pik}), the subchannels attaining positive transmit power will be allocated to $i_c^{\ast}$, and those with zero transmit power will not be allocated.
Thus, $\mathbf{K}_{i_c^{\ast}}$ is found by:
\begin{equation}
\mathbf{K}_{i_c^{\ast}}=\{ k | p_{i_c^{\ast},k}^{(\hat{k}')} > 0 \}.
\label{eq:Ki}
\end{equation}

The $g_{i_c,e_0,k}$ for $k (\in \mathbf{K}_{i_c}$) of the remaining users in set $\mathbf{U'_c} (= \mathbf{U_c} \setminus \{ i_c^{\ast} \} )$ is set to zero. This procedure is performed to avoid allocating the already allocated subchannels to any other CUE.
It can be observed from the Algorithm \ref{A1}'s subchannel allocation phase that this procedure is 
reciprocated until no CUE is rendered without a subchannel.

\subsection{Data Rate Adjustment Phase for CUEs}
\begin{figure}
\centering
\includegraphics[width=7.3cm]{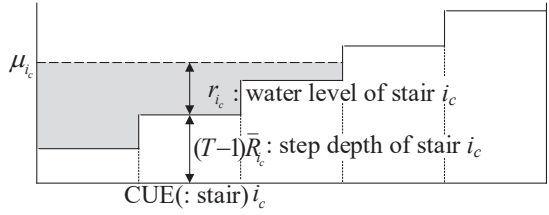}
\caption{The example of using a WF-based approach for data rate adjustment phase.}
\label{fig:WFConcept}
\end {figure}

In order to calculate $r_{i_c}$ (which satisfies (\ref{approximatedPF})), the $i_c$ (CUEs) are first sorted in increasing order of $(T-1)\bar {R}_{i_c}$. The detail of this procedure is provided in phase II of the Algorithm \ref{A1} i.e., the data rate adjustment phase of CUEs. Once the CUEs are properly ordered, the $r_{i_c}$ is can then be obtained using geometric WF technique \cite{PHe} as:
\begin{equation}
r_{i_c} = \left\{ \begin{array}{ll}
r_{\hat{i}_c} + (T-1)(\bar{R}_{\hat{i}_c} - \bar{R}_{i_c}), & \textrm{for } 1 \le i_c \le \hat{i}_c, \\
0, & \textrm{for } \hat{i}_c < i_c \le N_C,
\end{array} \right.
\label{eq:caseequation}
\end{equation}
where
\begin{equation}
\label{hati}
\hat{i_c}\!=\!\max\!\bigg \{ i_c \Bigg | \hat{r}_c - (T-1) \!\sum\!_{n=1}^{i_c-1}(\bar{R}_{i_c} - \bar{R}_{n}) \!>\! 0 \ \textrm{for }\forall i_c \!\!\in\!\! \mathbf{U_c} \bigg \}.
\end{equation}
An example of introducing WF based approach for obtaining ${r}_{i_c}$ is shown in Fig. \ref{fig:WFConcept} where the WF-based approach is applied to (\ref{approximatedPF}). Note that here we use `$i_c$' to represent the `index of the stair'. Likewise, `$r_{i_c}$' represents `water level poured into the stair $i_c$' and `$(T-1)\bar{R}_{i_c}$' represents the `step depth of stair $i_c$'.
Here, $\mu_{i_d,k'}$ represents the final water level in the D2D pair $i_d$'s water tank for $k'$.

In order to achieve proportional fairness, the PF scheduler can now adjust the user data rates by conforming the user MCS according to obtained $r_{i_c}$ (see \eqref{eq:caseequation} and \eqref{hati}).
In other words, the user's data rate is scaled down from $\hat{r}_{i_c}$ to $r_{i_c}$ by applying high-rate coding and low-order modulation schemes to users' information signals. For more details, the table given in \cite{ref:MCS} can be used as an example of how the MCS can be adjusted to achieve the given levels of user data rates.

\subsection{Subchannel Allocation Phase for D2D Pairs}
This phase also starts with finding all possible allocations for the D2D pairs under the adjacency constraint. This process is performed for all D2D pairs.
Since subchannels have already been assigned to CUEs, it is possible to determine which CUE may cause interference to a given D2D receiver on a particular subchannel.
By utilizing the already acquired information on $\bar{R}_{i_d}$ for all D2D pairs, the proposed scheme chooses the D2D pair $i_d^{\ast}$ (=$\arg\min_{i_d \in \mathbf{U_d}} (T-1)\bar{R}_{i_d}$).
in order to maximize the data rates of user $i_d^{\ast}$, the subchannel allocation problem can be formulated as:
\begin{eqnarray}
\max_{\{p_{i_d^{\ast},k}^{(k')}\}} & \sum_{k\in \mathbf{K}} B \cdot \log_{2} \Big ( {1+\frac{p_{i_d^{\ast},k}^{(k')}g_{i_d^{\ast T},i_d^{\ast R},k}}{N_0+p_{i_c(k)} \cdot g_{i_c(k),i_d^{\ast R},k}}} \Big ) \label{eq:water_hatri} \\
\textrm{subject to: } & \!\!\!\!\!\!\!\!\!\!\!\!\!\!\!\!\!\!\!\!\!\!\!\!\!\!\!\!\!\!\!\!\!\!\!\!\!\!\!\!\! p_{i_d^{\ast},k}^{(k')}+\frac{N_0+p_{i_c(k)} \cdot g_{i_c(k),i_d^{\ast R},k}}{g_{i_d^{\ast T},i_d^{\ast R},k}} \nonumber \\
&\!\!\!\!\!\!\!\!\!\!\!\!\!\ge p_{i_d^{\ast},k-1}^{(k')} + \frac{N_0+p_{i_c(k-1)} \cdot g_{i_c(k-1),i_d^{\ast R},k-1}}{g_{i_d^{\ast T},i_d^{\ast R},k-1}} \nonumber \\
&\ \ \ \ \ \ \ \ \ \ \ \ \ \ \textrm{for } \forall k\in \{ k'+1,\cdots,K\}; \label{eq:subject6} \\
& \!\!\!\!\!\!\!\!\!\!\!\!\!\!\!\!\!\!\!\!\!\!\!\!\!\!\!\!\!\!\!\!\!\!\!\!\!\!\!\!\!\!\!\!\!\!\!\!\!\!\!\!\!\!\!\!\!\!\!\! \sum_{k \in \mathbf{K}} p_{i_d^{\ast},k}^{(k')} \le \hat{P}_{i_d^{\ast}}, \label{eq:subject7} \\
& \!\!\!\!\!\!\!\!\!\!\!\!\!\!\!\!\!\!\!\!\!\!\!\!\!p_{i_d^{\ast},k}^{(k')} = 0 \textrm{ for } \forall k \!\! \in \!\! \{ 1,\cdots,k'-1\} \label{eq:subject8}
\end{eqnarray}
where $i_c(k)$ is the CUE $i_c$ which already occupies subchannel $k$, $p_{i_c(k)}$ is the transmit power of CUE $i_c(k)$ and $g_{i_c(k),i_d^\ast,k}$ is the channel gain between CUE $i_c(k)$ and D2D receiver of the pair $i_d^{\ast}$. Likewise, the $p_{i_c(k)} \cdot g_{i_c(k),i_d^\ast,k}$ and $\hat{P}_{i_d^{\ast}}$ are the interference caused by CUE $i_c(k)$ to the D2D receiver in the pair $i_d^\ast$ and the system defined maximum transmit power of $i_d^{\ast}$'s, respectively.

By applying the modified cap-limited water-filling approach \cite{LZhuang}, the optimal $\{ p_{i_d^{\ast},k}^{(k')} \}$ in the proposed algorithm can be calculated as:
\begin{equation}
p_{i_d^{\ast},k}^{(k')} = \left\{ \begin{array}{ll}
\big[ \mu_{i_d^\ast,k'} - \frac{N_0+p_{i_c(k)} \cdot g_{i_c(k),i_d^{\ast R},k}}{g_{i_d^{\ast T},i_d^{\ast R},k}} \big] ^{+}, & \textrm{for } k=k',\\
\big[ \mu_{i_d^{\ast},k'} - p_{i_d^{\ast},k-1}^{(k')} \\
- \frac{N_0+p_{i_c(k-1)} \cdot g_{i_c(k-1),i_d^{\ast R},k-1}}{g_{i_d^{\ast T},i_d^{\ast R},k-1}} \\
+ \frac{N_0+p_{i_c(k)} \cdot g_{i_c(k),i_d^{\ast R},k}}{g_{i_d^{\ast T},i_d^{\ast R},k}} \big] ^{+}, & \\
& \!\!\!\!\!\!\!\!\!\!\!\!\!\!\! \textrm{for } k' < k \le K, \\
0, & \textrm{otherwise,}
\end{array} \right.
\label{eq:pikd2d}
\end{equation}
where $[\phi]^{+} = \max \{ 0, \phi \}$, and $\mu_{i_d^\ast,k'}$ is the level of water for D2D pairs in case $k'$ for which (\ref{eq:subject4}) is satisfied with equality.
Furthermore, the $\hat{r}_{i_d^{\ast}}^{(k')}$ can also be computed as:
\begin{equation}
\hat{r}_{i_d^{\ast}}^{(k')} = \sum_{k\in \mathbf{K}} B \cdot \log_{2} \Big ( {1+\frac{p_{i_d^{\ast},k}^{(k')}g_{i_d^{\ast T},i_d^{\ast R},k}}{N_0+p_{i_c(k)} \cdot g_{i_c(k),i_d^{\ast R},k}}} \Big ).
\label{eq:riastkd2d}
\end{equation}
Then for the case $\hat{k}'$ ($= \arg \max_{\{k'\}} \hat{r}_{i_d^{\ast}}^{(k')}$), the $\hat{r}_{i_d^{\ast}}$ can decided by:
\begin{equation}
\hat{r}_{i_d^{\ast}} = \max_{\{k'\}} \hat{r}_{i_d^{\ast}}^{(k')}.
\label{eq:maxrd2d}
\end{equation}
After this, we determine the set $\mathbf{K}_{i_d^{\ast}}$ of subchannels, which will be allocated to D2D pair $i_d^{\ast}$ for $\hat{k}'$.
From (\ref{eq:pikd2d}), the subchannels having a positive transmit power value will be allocated to $i_d^{\ast}$, and those with zero transmit power will not be allocated.
Finally, for the subchannel allocation case $\hat{k}'$ the subchannels set $\mathbf{K}_{i_d^{\ast}}$ of positive power is allocated to D2D pair $i_d$ such that:
\begin{equation}
\mathbf{K}_{i_d^{\ast}}=\{ k | p_{i_d^{\ast},k}^{(\hat{k}')} > 0 \}.
\label{eq:Kid2d}
\end{equation}
The $g_{i_d^T,i_d^R,k}$ for $k (\in \mathbf{K}_{i_d}$) of the remaining users in the set $\mathbf{U'_d} (= \mathbf{U_d} \setminus \{ i_d^{\ast} \} )$ is set to zero. This procedure is performed to avoid the allocation of already allocated subchannels to any other D2D pair.
It can be observed from the procedure of the subchannel allocation phase in Algorithm \ref{A1} that this process is repeatedly performed until no D2D pair is rendered without a subchannel.

\subsection{Data Rate Adjustment Phase for D2D Pairs}
\textbf{In order to calculate $r_{i_d}$ that satisfies (\ref{approximatedPF}), this phase of our proposed algorithm first sorts the $i_d$ in increasing order of $(T-1)\bar {R}_{i_d}$.
Then according to the geometric WF approach \cite{PHe}, $r_{i_d}$ is obtained as:
\begin{equation}
r_{i_d} = \left\{ \begin{array}{ll}
r_{\hat{i}_d} + (T-1)(\bar{R}_{\hat{i}_d} - \bar{R}_{i_d}), & \textrm{for } 1 \le i_d \le \hat{i}_d, \\
0, & \textrm{for } \hat{i}_d < i_d \le N_D,
\end{array} \right.
\label{eq:caseequationd2d}
\end{equation}
where
\begin{equation}
\label{hatid2d}
\hat{i_d}\!=\!\max\! \bigg \{ i_d \Bigg | \hat{r}_d - (T-1)\!\! \sum\!_{n=1}^{i_d-1}(\bar{R}_{i_d} - \bar{R}_{n}) \!>\! 0 \ \textrm{for }\forall i_d \!\in\! \mathbf{U_d}\! \bigg \}.
\end{equation}
Based on this $r_{i_d}$, the PF scheduler may tune the user data rates by adjusting their MCS to achieve desired proportional fairness.}

\section{Performance Evaluation}\label{sec:pe}

To thoroughly evaluate the performance, We have executed system-level simulations for the proposed and optimal PF scheduling schemes using a homegrown C/C++ programming-based simulator.
In Section \ref{sec:pe}, we call the Proposed Heuristic PF Scheduler as PHPFS and the optimal PF scheduling as O-PF.
As the performance metric, the simulation prints out the computing complexity and the logarithmic sum of the average of achievable user data rates.
A frequency division duplex (FDD)-based LTE system has been designed in the simulation.
The system consists of 19 cells that follow the layout option 3 (with 57 wrap-around hexagonal sectors) given by \cite{TR36843}.
$N_C$ CUEs and $N_D$ D2D pairs are deployed uniformly in each of the 19 cells.
UEs' mobility has been implemented with the random-walk model.
In the model, UEs' speeds and directions are uniformly decided between [0, 10] (m/s) and between [0, 2$\pi$], respectively.
When a UE approaches the borderline of the concerned cell or when the flight time expires that is uniformly chosen in [10, 30] (sec), the speed and direction are newly decided.
A fully-buffered traffic model has been applied to all UEs to find the upper limit of users' data rates. 
Single-input and single-output (SISO) channels are considered for wireless channel modelling. The path loss \cite{ref:itu-r}, large-scale fading (shadowing) \cite{ref:shadowing} and small-scale fading \cite{ref:multipath} are also taken into account.
In order to compute the frequency efficiency (bps/Hz) at a certain SINR, the MCS table provided in \cite{ref:MCS} is used.
The numerical values of the simulation parameters are given in Table \ref{tb:param}.
\begin{table}
\centering
\caption{Various Parameters and Their Values Used in Simulation.}
\label{tb:param}
\begin{tabular}{|c|c|}
\hline
Parameters Name &\,\,\,\,\,\,\,\,\,\,\,\, Value \,\,\,\,\,\,\,\,\,\,\,\,\\
\hline
\hline
eNB receiving antenna gain & 15.0 dB \\
\hline
UE receiving antenna gain& 4.0 dB \\
\hline
Center frequency of carrier & 2.0 GHz \\
\hline
Decorrelation length for shadowing & 50 m \\
\hline
Noise figure & 5.0 dB\\
\hline
Power density of noise & -174 dBm/Hz \\ 
\hline
Bandwidth of subchannel & 180 KHz \\
\hline
Time interval between transmissions & 1 msec \\
\hline
Transmit power of UE & 23 dBm \\ 
\hline
\end{tabular}
\end{table}

\subsection{Computing Complexity}

The complexity under the worst case of schedulers is given in Table \ref{tb:complexity}.
\begin{table}
\centering
\caption{Computing Complexities of Optimal PF and PHPFS.}
\label{tb:complexity}
\begin{tabular}{|c|c|}
\hline
Scheduler & Computing Complexity\\
\hline
\hline
Optimal PF scheduling & $(N_C+N_D)^{K}$ \\
\hline
& $\{K\log(K)+5K\} $\\
Proposed scheduling algorithm &$\times \{N_C^2(N_C+1)/2 $\\
&$+ N_D^2(N_D+1)/2\} \times M$ \\
\hline
\end{tabular}
\end{table}
It can be observed from Table \ref{tb:complexity} that the computing complexity of PHPFS rises linearly with $K\log(K)+5K$.
Contrarily, the complexity of the optimal PF scheduler exponentially increases with $K$.
This is due to the fact that the optimal PF scheduling searches for all possibilities that $N$ users are successfully allocated to $K$ carriers.
If we apply the complexity analysis covered in \cite{LZhuang} and \cite{PHe} on PHPFS, we find that in the resource allocation phase, every iteration has the complexity of $K\log(K)+5K$, and for data rate adjustment phase it is $N^2(N+1)/2$.

\begin{figure}
\centering
\subfigure[When the value of $K$ is set to 3, 5, and 10 (small values of $K$).]{
\label{fig:complexity_smallK}
\includegraphics[width=8.8cm]{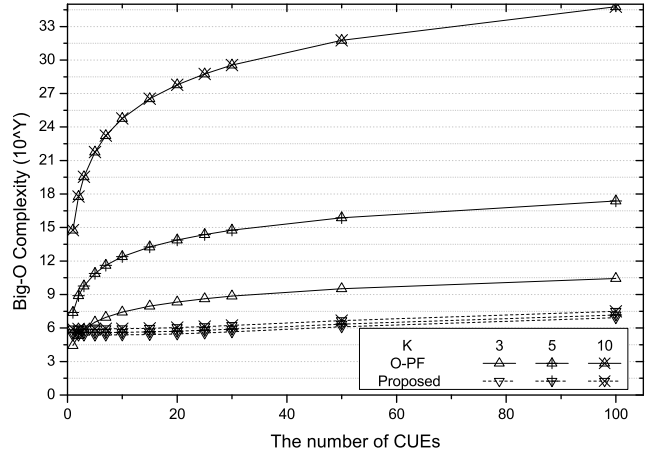}}
\subfigure[When the value of $K$ is set to 25, 50, and 75 (large values of $K$).]{
\label{fig:complexity_largeK}
\includegraphics[width=8.8cm]{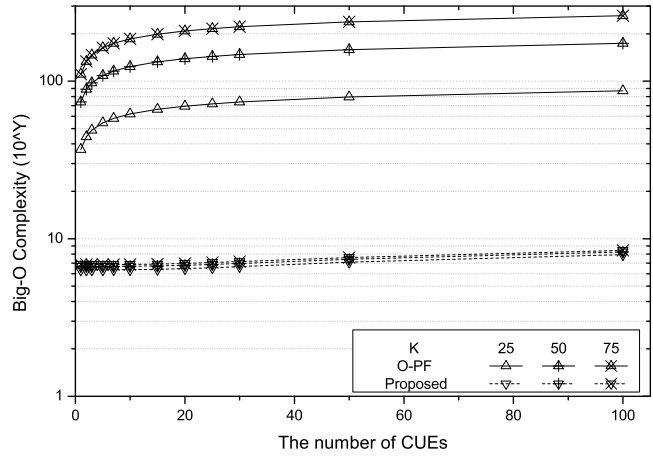}}
\caption{The computational complexity comparison of both O-PF and PHPFS schemes.}
 \label{fig:complexity}
\end{figure}

In order to examine the computing complexity, we start with smaller values of $K$, i.e., $K=$ 3, 5, and 10 for both PHPFS and optimal PF scheduling schemes.
Since the LTE system splits 5 MHz bands into 25 subchannels and 10 MHz bands into 50 subchannels \cite{ref:LTEtheUMTS}, the varying values of $K$ ($=$ 25, 50, and 75) are considered in our simulations.
Fig. \ref{fig:complexity} shows the computing complexity for $K$ carriers and 30 D2D pairs.
For smaller values of $K$(=3, 5, and 10), Fig. \ref{fig:complexity_smallK} shows that increasing $K$ raises the computational complexity of O-PF.
However, the computing complexity of PHPFS is nearly constant.
This is because the computational complexity of the PHPFS is almost linearly increasing with $(N_C)$.
When $K=3$ and $N_C < 5$, PHPFS has larger computing complexity compared to O-PF.
On the other hand, for practically used values ($N_C \ge 5$ and $K\ge3$), the computing complexity of PHPFS is low.

\textbf{
Fig. \ref{fig:complexity_largeK} shows the computational complexity for $K=$ 25, 50, and 75 for all values of $N_C$.
It can be observed that as the value of $K$ increases, the difference between the computational complexity of PHPFS and the O-PF schedule also increases.
When $K=75$ and $N_D=100$, the O-PF has about $10^{200}$ times higher computational complexity than the PHPFS.
Even when $K=25$ and $N_D=3$, it can be verified from Fig. \ref{fig:complexity_largeK} that the PHPFS scheme significantly reduces the computational complexity of O-PF as much as $10^{30}$.
In real-life practical environments, every subchannel experiences independent small-scale fading, completely different from any other subchannel used in the network.
Therefore, the data rate achieved by user $i$ on a carrier $k$ is not usually the same as that achieved by using another carrier $k'$.
However, in our simulation environment, we assume that the small-scale fading of every channel is the same (i.e., flat small-scale fading) for $K\ge25$.
The above-mentioned assumption will help to finish the simulations in a reasonable time.}

\subsection{Logarithmic Sum of Achievable User Data Rates with Given $N_D$}

In this section, we evaluate the performance of both PHPFS and O-PF schemes regarding the logarithmic sum of average user data rates.
The performance of PHPFS is evaluated using one, two, and three iterations (i.e., $M=1, 2, 3$) in order to reduce the simulation time. 
\begin{figure}
\centering
\subfigure[$K=5$.]{
\label{fig:logsum_k5}
\includegraphics[width=8.8cm]{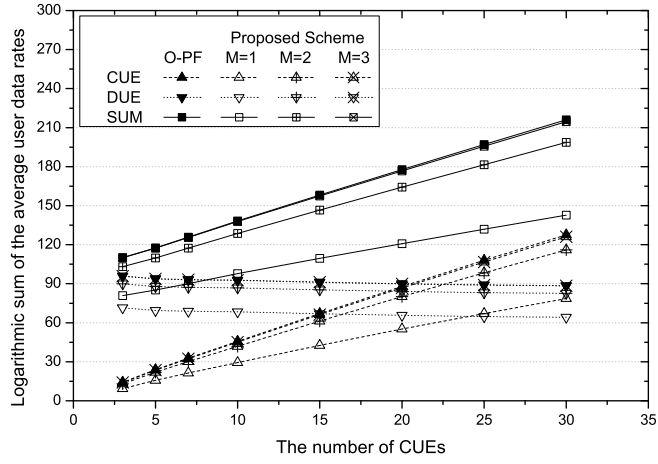}}
\subfigure[$K=10$.]{
\label{fig:logsum_k10}
\includegraphics[width=8.8cm]{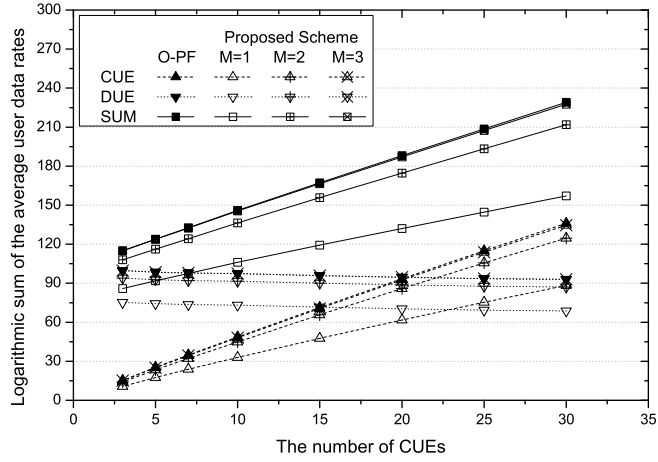}}
\caption{For $K=$ 5 and 10, the logarithmic sum of the average user data rates when $N_D=$20.}
\label{fig:graphlogsum}
\end{figure}
The results are illustrated in Figure \ref{fig:graphlogsum}, and it can be observed that as $N_C$ increases, for both PHPFS and O-PF, the logarithmic sum for CUEs' average data rates rises while that for DUEs is reduced.
This is because of the increased interference at DUEs introduced by the large number of $N_C$.
It is intuitively known that O-PF can achieve the best logarithmic sum of average data rates.
However, PHPFS also achieves the logarithmic sum of average data rates nearly identical to O-PF for CUEs.

Furthermore, it can also be observed that with the increasing number of D2D pairs, the average throughput per DUE is reduced.
This is because the available number of subchannels for any specific D2D pair decreases as the number of active D2D pairs in the network increases. 
Similarly, it can also be observed that the logarithmic sum of average achievable data rates of DUEs in the PHPFS scheme is reduced.
The PF metric is not completely considered during the subchannel scheduling.
Therefore, the logarithmic sum for DUEs achieved by the PHPFS scheme is lower than that of the O-PF scheme.
It has also been observed that for a small number of D2D pairs, the performance of PHPFS is close to that of O-PF scheduling. Furthermore, a similar trend can also be observed in Fig. \ref{fig:logsum_k10}.

\begin{figure}
\centering
\subfigure[$K=5$.]{
\label{fig:datarate_k5}
\includegraphics[width=8.8cm]{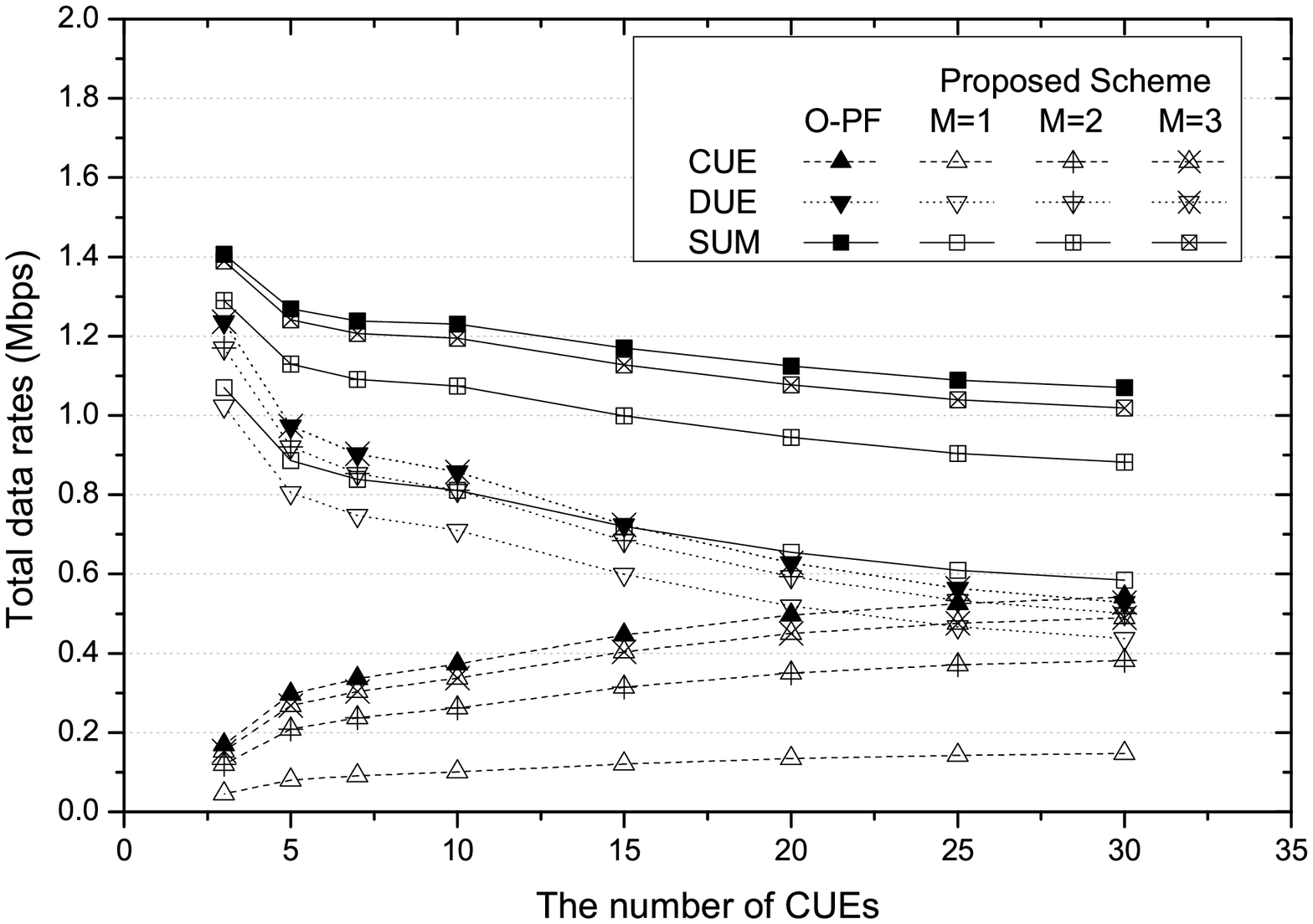}}
\subfigure[$K=10$.]{
\label{fig:datarate_k10}
\includegraphics[width=8.8cm]{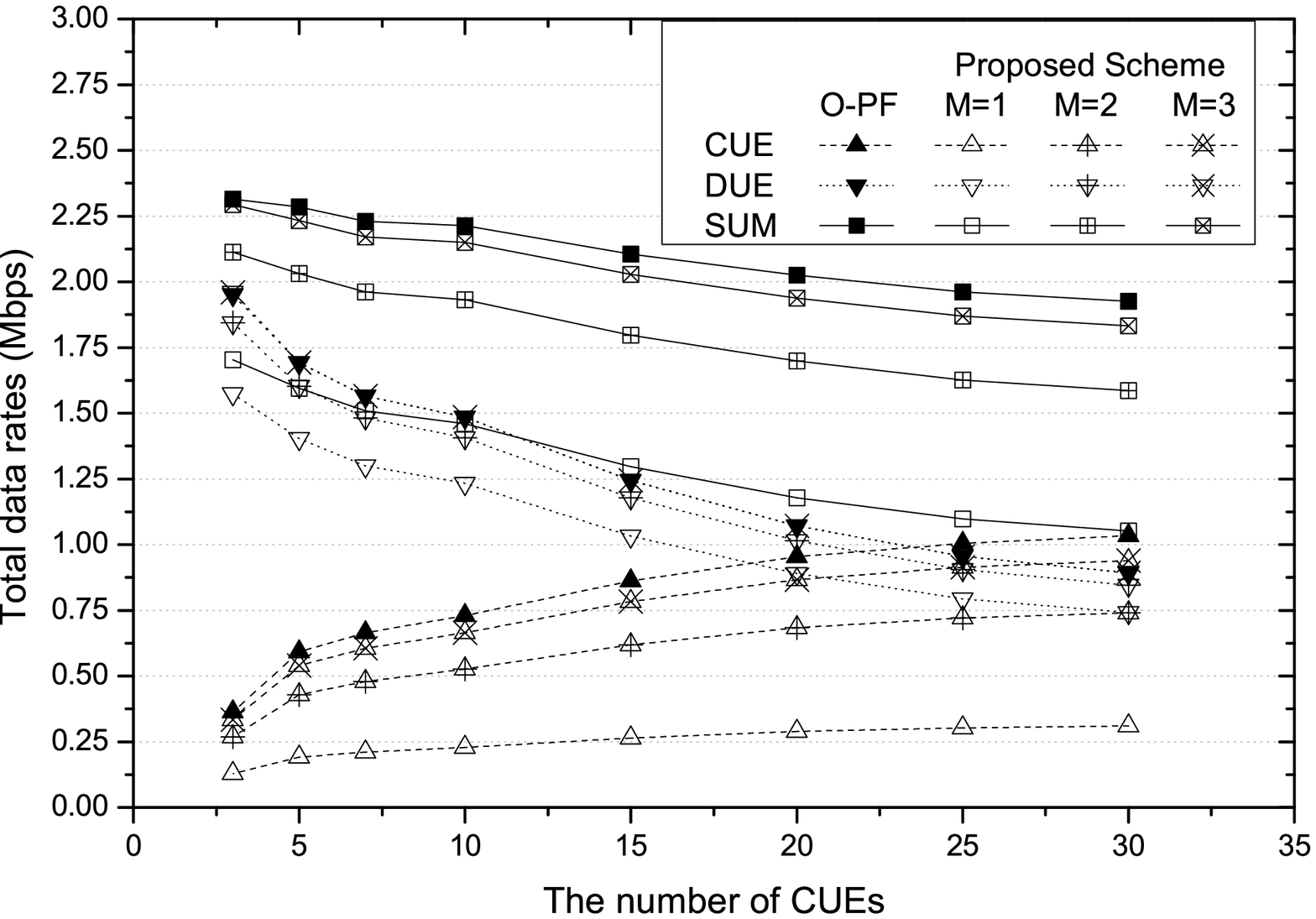}}
\caption{For $K=$ 5 and 10, the total user data rates when $N_D=$20.}
\label{fig:graphdatarate}
\end{figure}

For $K=5$ and $K=10$, the individual data rates of both DUEs, CUEs, and their joint data rate are also evaluated, and their corresponding results are provided in Figure \ref{fig:graphdatarate}.
Figure \ref{fig:datarate_k5} illustrates that the CUEs' best throughput is ranged within [0.05, 1.5] Mbps for PHPFS.
On the other hand, in the O-PF scheduler, the best throughput for CUEs' is ranged within [0.2, 1.7] Mbps, which is obviously better than that of PHPFS.
This is due to the fact that the O-PF scheme considers the data rates of UEs before assigning the subchannels, which, as a result, offer better opportunities for getting radio resources to CUEs and DUEs with better data rates.
Note that the difference between the three iterations of PHPFS and O-PF is not very significant.
Fig. \ref{fig:datarate_k10} shows that the trend of the data rates is nearly the same for $K=5$ and $K=10$.

\begin{figure}
\centering
\subfigure[$K=25$.]{
\label{fig:logsum_k25}
\includegraphics[width=8.8cm]{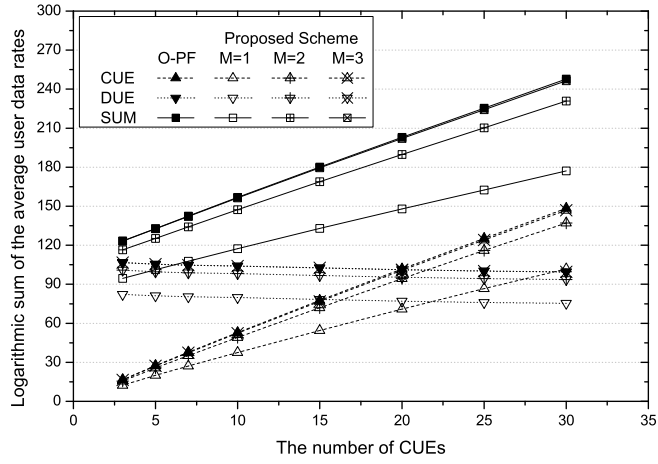}}
\subfigure[$K=50$.]{
\label{fig:logsum_k50}
\includegraphics[width=8.8cm]{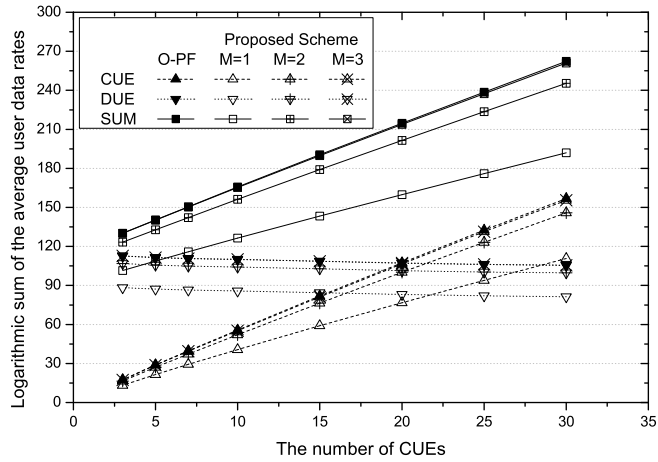}}
\caption{For $K=$ 25 and 50, the logarithmic sum of the average user data rates when $N_D=$20.}
\label{fig:graphmultilogsum}
\end{figure}

The logarithmic sum of CUEs and DUEs' average data rates and their sum for $K=$ 25 and 50 are shown in Figs. \ref{fig:logsum_k25} and \ref{fig:logsum_k50}, respectively.
Similar to the result shown in Fig. \ref{fig:graphlogsum}, increasing the value of $N_D$ increases the logarithmic sum.
The logarithmic sums for $K=$ 25 and 50 are increased from that for $K=$ 5 and 10 because the increased number of subchannels are allocated to a D2D pair with increasing $K$.
The rate of increase in the logarithmic sum for DUEs is decreased in the PHPFS as the number of D2D pairs increases.

\begin{figure}
\centering
\subfigure[$K=25$.]{
\label{fig:datarate_k25}
\includegraphics[width=8.8cm]{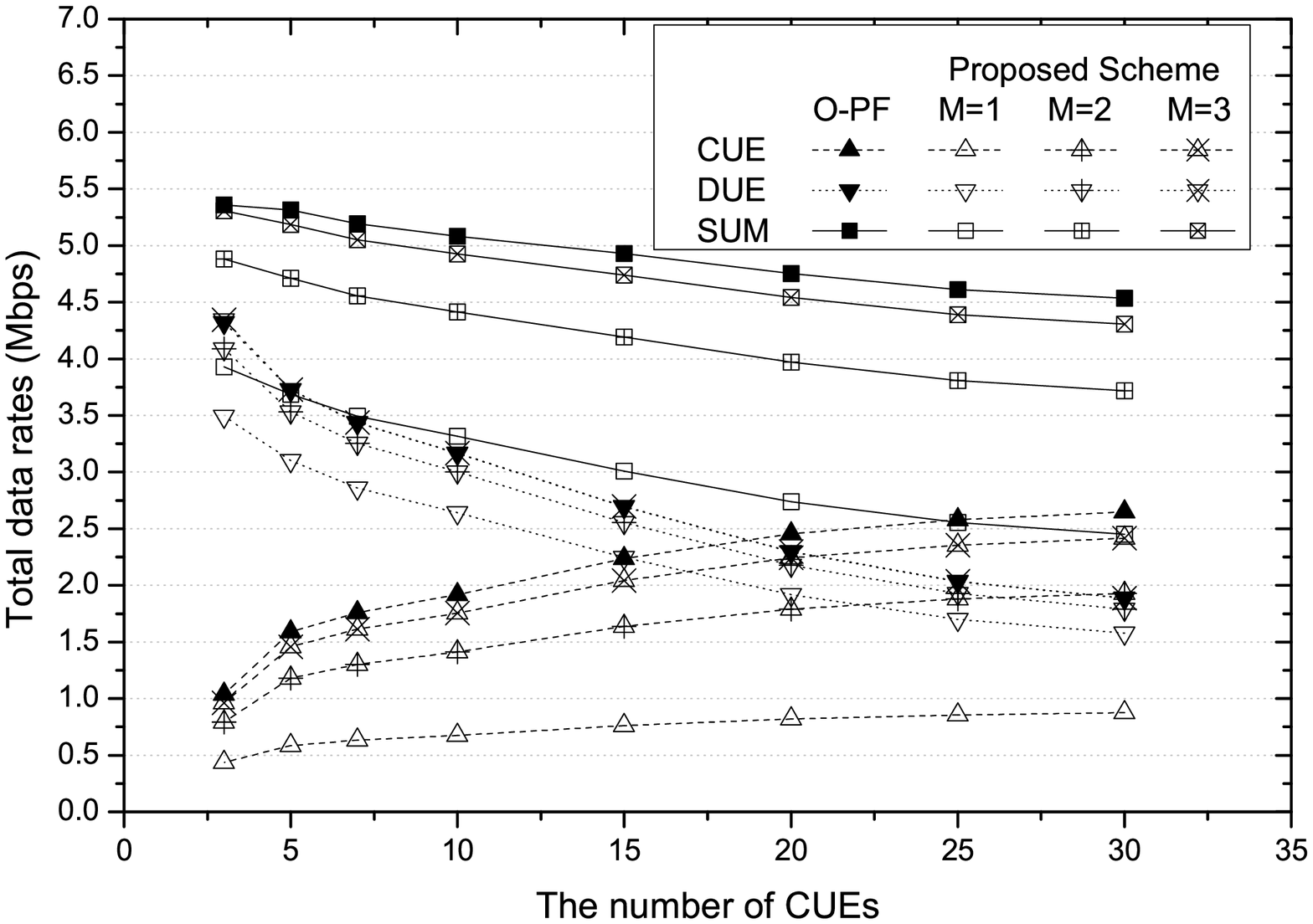}}
\subfigure[$K=50$.]{
\label{fig:datarate_k50}
\includegraphics[width=8.8cm]{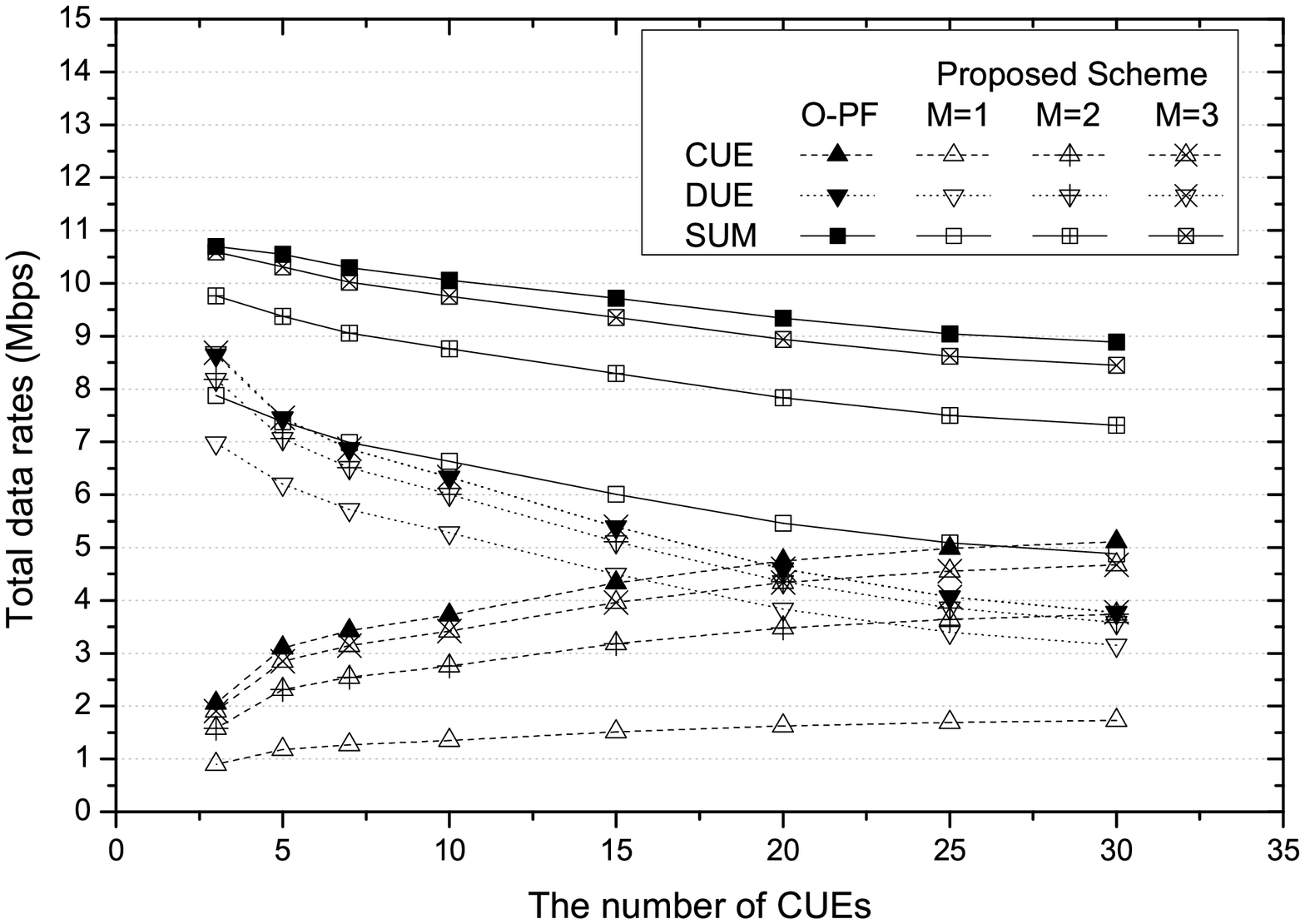}}
\caption{For $K=$ 25 and 50, the total user data rates when $N_D=$20.}
\label{fig:graphmultidatarate}
\end{figure}

The achievable data rates of CUEs, DUEs, and their sum for both O-PF and PHPFS when $K=$ 25 and 50 are depicted in Figure \ref{fig:graphmultidatarate}.
It can be observed that the O-PF has higher data rates for CUEs than that of PHPFS in all of the cases ($K = $ 25 and 50, $N_C = $ 3 to 30).
The reason behind this is similar to what is explained earlier, i.e., that the optimal scheme maximizes the average data rates of the CUEs'. However, given the computational complexity gains of PHPFS, the difference in the throughput performance of both schemes is negligible.

\subsection{The Logarithmic Sum of Both Total and Average Achievable User Data Rates When $N_C$ is Fixed }

\begin{figure}
\centering
\subfigure[$K=5$.]{
\label{fig:logsum_k5_ND}
\includegraphics[width=8.8cm]{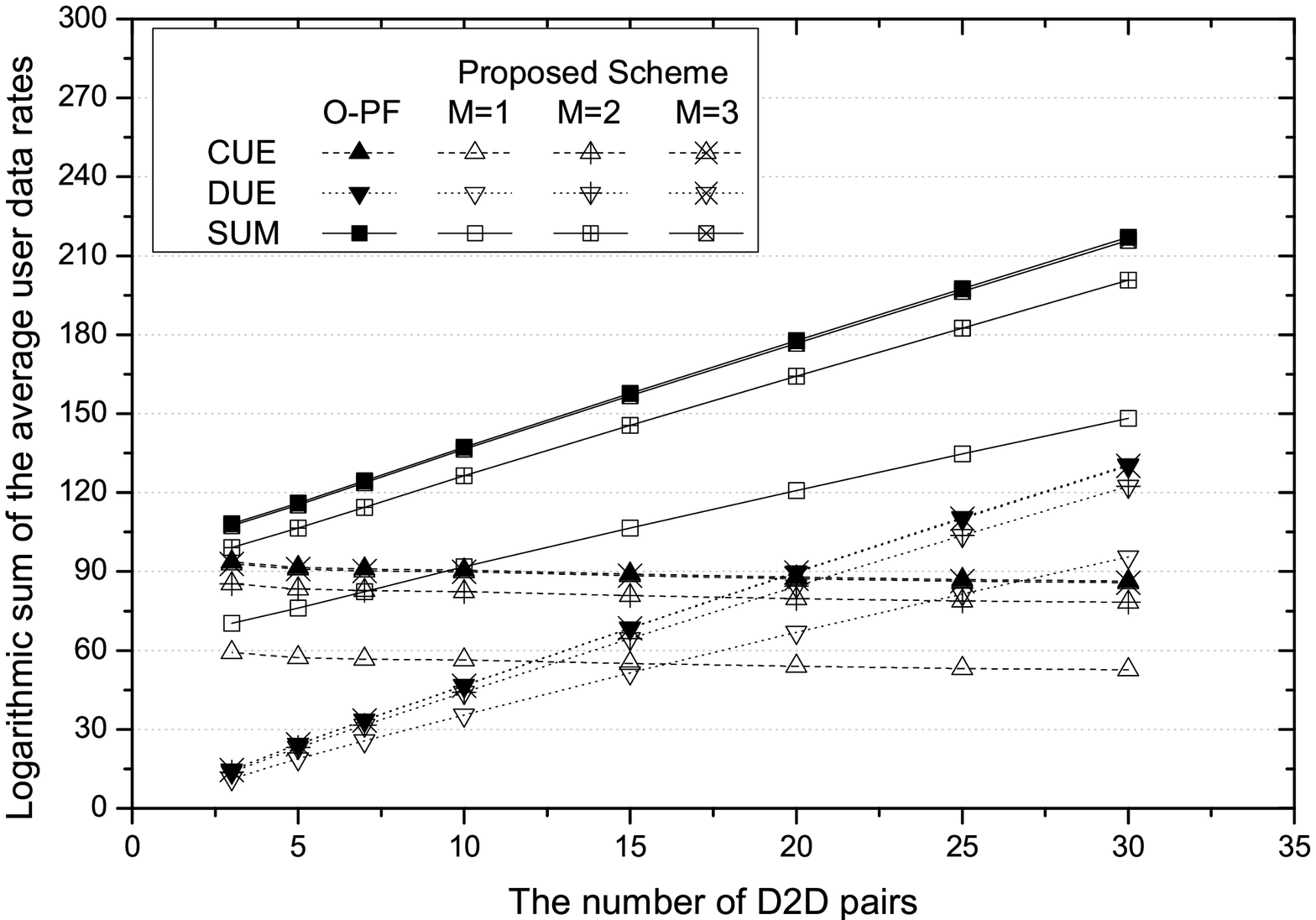}}
\subfigure[$K=50$.]{
\label{fig:logsum_k50_ND}
\includegraphics[width=8.8cm]{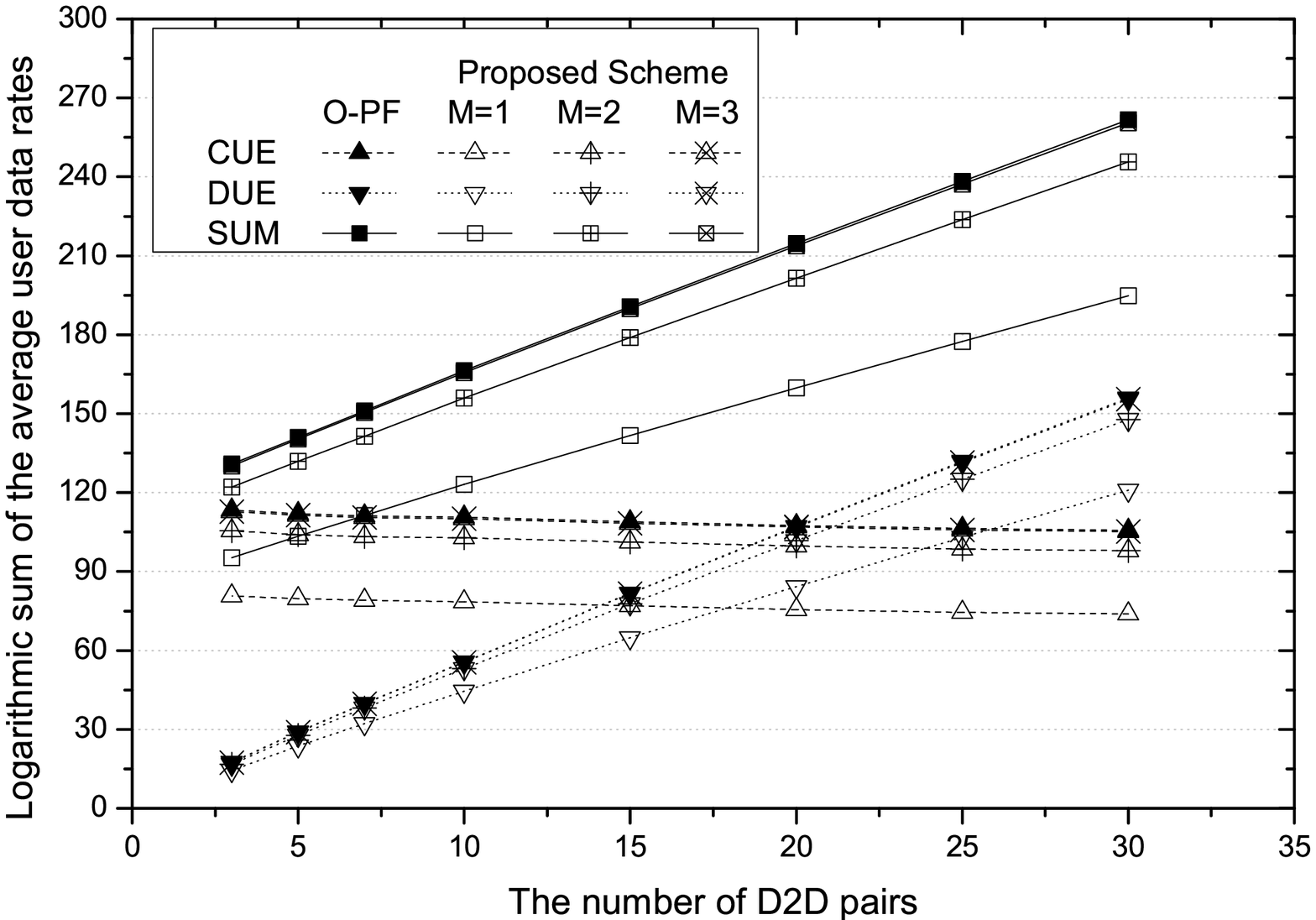}}
\caption{For $K=$ 5 and 50, the logarithmic sum of the users' average data rates with varying numbers of $N_D$ (where $N_C=$20).}
\label{fig:graphlogsum_ND}
\end{figure}

Fig. \ref{fig:graphlogsum_ND} shows the logarithmic sum of CUEs with varying values of $N_D$. In this case, 20 CUEs are considered, and the value is set to $K=$ 5 and 50.
It is shown that with the increasing value of $N_D$, the logarithmic sum increases.
It can also be observed that for $K=$50, the logarithmic sum is higher than that of $K=$5. It is because of the fact that the larger value of $K$ results in more subchannels allocated to each DUE. 
For O-PF, it is intuitive that it will always maximize the logarithmic sum for CUEs, DUEs, and the combined.

\begin{figure}
\centering
\subfigure[$K=5$.]{
\label{fig:datarate_k5_ND}
\includegraphics[width=8.8cm]{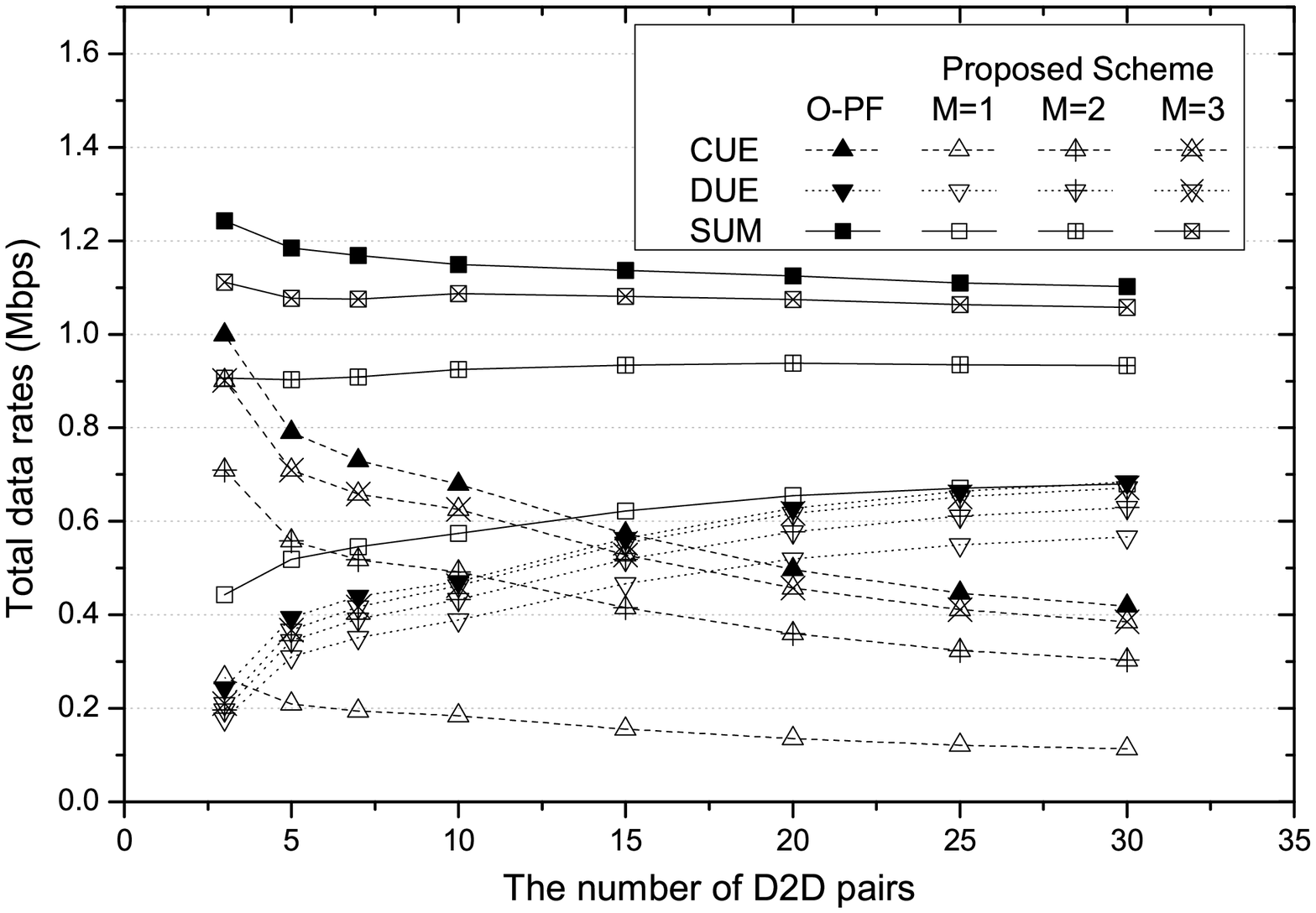}}
\subfigure[$K=50$.]{
\label{fig:datarate_k50_ND}
\includegraphics[width=8.8cm]{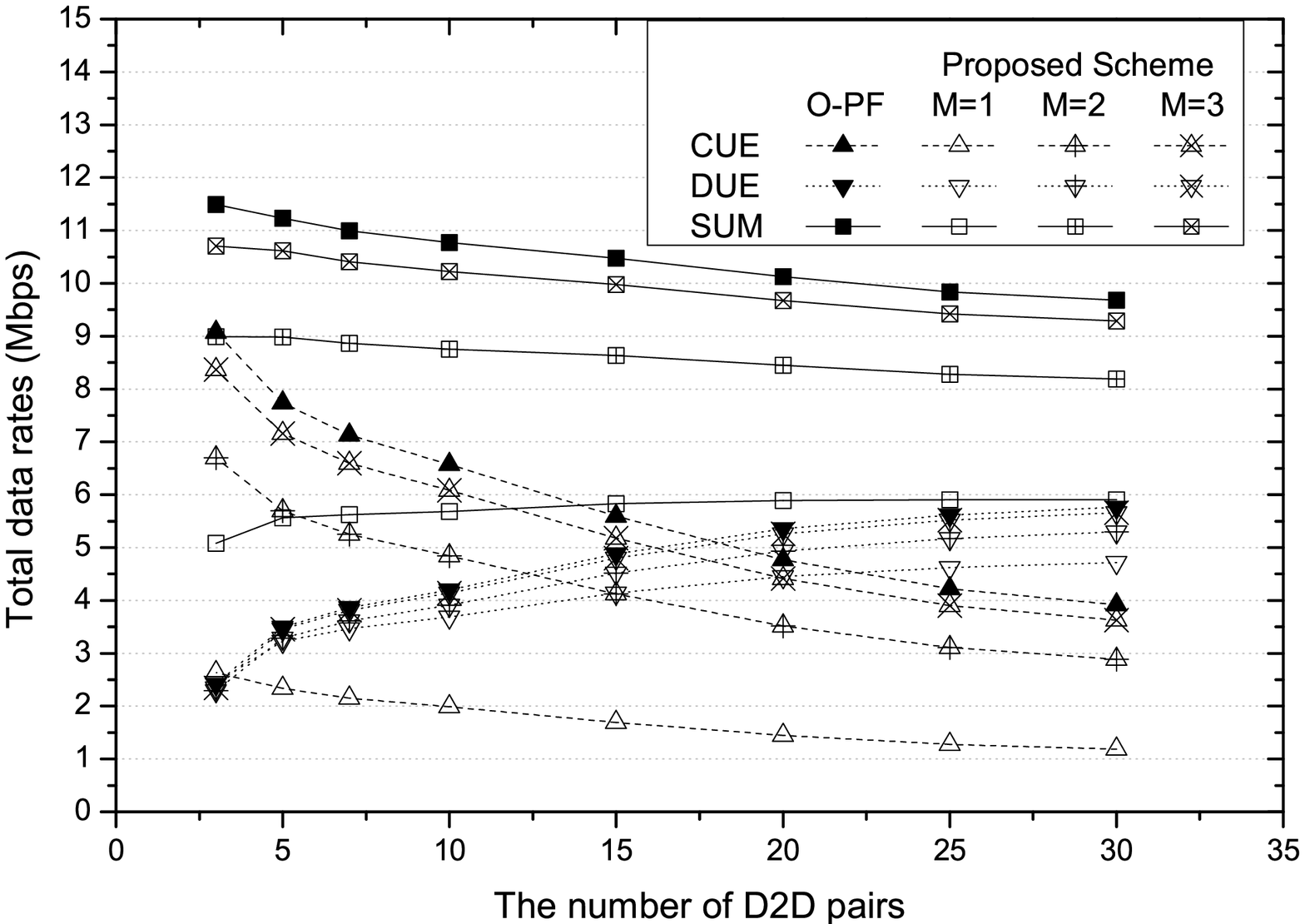}}
\caption{For $K=$ 5 and 50, the total user data rates with different numbers of $N_D$ (where $N_C=$20).}
\label{fig:graphdatarate_ND}
\end{figure}

For $K=$ 5 and 50, the overall data rates of both O-PF and PHPFS schemes with varying numbers of$N_Ds$ are depicted in Fig. \ref{fig:graphdatarate_ND}. Similar to Fig. \ref{fig:graphlogsum_ND}, the values of CUEs, in this case, are also fixed to 20.
Figure \ref{fig:graphdatarate_ND} shows that O-PF provides the best data rates for CUEs and D2D pairs.
In addition, we can also observe that increasing the number of D2D pairs in the network also decreases the CUEs data rates. It is because of the increased interference in the network generated by growing numbers of D2D pairs.
Contrarily, as the number of D2D pairs in the network grows, the available subchannels are highly reused, and an overall improved DUE data rate can be observed.
Since the increased data rates of DUEs can compensate for the reduced data rates of CUEs, the combined sum of data rates for both CUEs and DUEs remains almost the same.

\section{Conclusion}
This paper deals with the PF scheduling for D2D systems that adopts SC-FDMA.
The discussion starts with an introduction to D2D communication underlying cellular systems and the constraints it must conform to use SC-FDMA.
It has been argued that optimal PF scheduling in SC-FDMA uses brute-force search for all possible cases, which increases the computational complexity given the large number of users and sub-channels.
To reduce computing complexity, a heuristic algorithm is proposed for PF scheduling in SC-FDMA-based systems.
The well-known water-filling technique has been used for this purpose.

This paper has evaluated the computing complexity and user data rates.
From the perspective of computation complexity, the proposed scheduling has performed considerably better than the PF scheduling.
Meanwhile, the PF scheduling has performed better overall data rate than the proposed algorithm.
From the perspective of the logarithmic sum, both PF and proposed scheduling schemes perform the same for a small number of D2D pairs.
In conclusion, the proposed heuristic algorithm will perform reasonably but with low computational overhead.


\end{document}